\documentclass[12pt,journal,draftcls,letterpaper,onecolumn]{IEEEtran}
\usepackage{amscd}
\usepackage{amsfonts,latexsym,subfigure,graphicx,amsmath,algpseudocode,booktabs, balance}
\usepackage{algorithm}
\usepackage[usenames,dvipsnames]{pstricks}
\usepackage{epsfig}
\usepackage{pst-grad} 
\usepackage{pst-plot} 
\usepackage{multirow}
\usepackage{amsmath} 
\usepackage{amssymb}  

\newtheorem{defn}{\textsc{Definition}}
\newtheorem{exam}{\textsc{Example}}

\begin{document}
\title{HeteSim: A General Framework for Relevance Measure in Heterogeneous Networks}
\author{Chuan~Shi,~\IEEEmembership{Member,~IEEE,}
        Xiangnan~Kong,~\IEEEmembership{}
        Yue~Huang,~\IEEEmembership{}
        Philip~S.~Yu,~\IEEEmembership{Fellow,~IEEE,}
        and~Bin~Wu,~\IEEEmembership{Member,~IEEE}
\IEEEcompsocitemizethanks{\IEEEcompsocthanksitem C. Shi, Y. Huang, and B. Wu are with Beijing University of Posts and Telecommunications, Beijing, China.\protect\\
E-mail: shichuan@bupt.edu.cn, ymoon.huang@gmail.com, wubin@bupt.edu.cn.
\IEEEcompsocthanksitem X.N. Kong and P.S. Yu are with University of Illinois at Chicago, IL, USA.
E-mail: kongxn@gmail.com, psyu@uic.edu.
}
\thanks{}}
\markboth{IEEE Transactions on Knowledge and Data Engineering,,~Vol.~6, No.~1, January~2007}{Shell \MakeLowercase{\textit{C. Shi et al.}}: HeteSim: A General Framework for Relevance
Measure in Heterogeneous Networks}

\maketitle

\begin{abstract}
Similarity search is an important function in many applications, which usually focuses on measuring the
similarity between objects with the same type. However, in many
scenarios, we need to measure the relatedness between
objects with different types. With the surge
of study on heterogeneous networks, the relevance measure on
objects with different types becomes increasingly important. In this
paper, we study the relevance search problem in heterogeneous
networks, where the task is to measure the relatedness of
heterogeneous objects (including objects with the same type or
different types). A novel measure HeteSim is proposed, which has
the following attributes: (1) a uniform measure: it can measure the relatedness of objects
with the same or different types in a uniform framework; (2) a path-constrained measure: the
relatedness of object pairs are defined based on the search path
that connect two objects through following a sequence of node types;
(3) a semi-metric measure: HeteSim has some good properties (e.g.,
self-maximum and symmetric), that are crucial to many data mining tasks. Moreover, we analyze the computation characteristics of HeteSim and propose the corresponding quick computation strategies.
Empirical studies show that HeteSim can effectively and efficiently evaluate the
relatedness of heterogeneous objects.
\end{abstract}


\section{Introduction}
%
%
%
%
\PARstart{S}{imilarity} search is an important task in a wide range of
applications, such as web search \cite{PBMW98} and product
recommendations \cite{KMMHG97}. The key of similarity search is
similarity measure, which evaluates the similarity of object pairs.
Similarity measure has been extensively studied for traditional
categorical and numerical data types, such as Jaccard coefficient
and cosine similarity. There are also a few studies on leveraging
link information in networks to measure the node similarity, such as
Personalized PageRank \cite{JW03}, SimRank \cite{JW02}, and PathSim
\cite{SHYYW11}. Conventional study on similarity measure focuses on
objects with same type. That is, the objects being measured are of the
same type, such as ``document-to-document", ``webpage-to-webpage"
and ``user-to-user". There is seldom research on similarity measure
on objects with different types. That is, the objects being measured are
of different types, such as ``author-to-conference'' and
``user-to-movie''. It is reasonable. The similarity of
objects with different types is a little against our common sense.
Moreover, different from the similarity of objects with same type, which
can be measured on homogeneous situation (e.g., the same
feature space or homogeneous link structure), it is even hard to
define the similarity of objects with different types.

However, the similarity of objects with different types is not only meaningful but also useful in some scenarios.
For example, the author J. F. Naughton is more relevant to SIGMOD
than KDD. A teenager may like the movie ``Harry Potter" more than
``The Shawshank Redemption". Moreover, the similarity measure of
objects with different types are needed in many applications. For
example, in a recommendation system, we need to know the relatedness
between users and movies to make accurate recommendations. In an
automatic profile extraction application, we need to measure the relatedness of
objects with different types, such as authors and conferences,
conferences and organizations etc. Particularly, with the advent of
study on heterogeneous information networks \cite{SHYYW11,SYH09}, it is not only increasingly important but also feasible to study the relatedness among objects with different types.
Heterogeneous information networks are the logical networks
involving multiple-typed objects and multiple-typed links denoting
different relations \cite{Han09}. For example, a bibliographic network
includes authors, papers, conferences, terms and their links
representing their relations. It is clear that heterogeneous information
networks are ubiquitous and form a critical component of modern
information infrastructure \cite{Han09}. So it is essential to provide a relevance
search function on objects with different types in such networks, which
is the base of many applications. Since objects with different types
coexist in the same network, their relevance measure is possible through link structure.

%

\begin{figure}[t]
    \begin{center}
    \subfigure[Symmetric measure]
    {\includegraphics[scale=0.25]{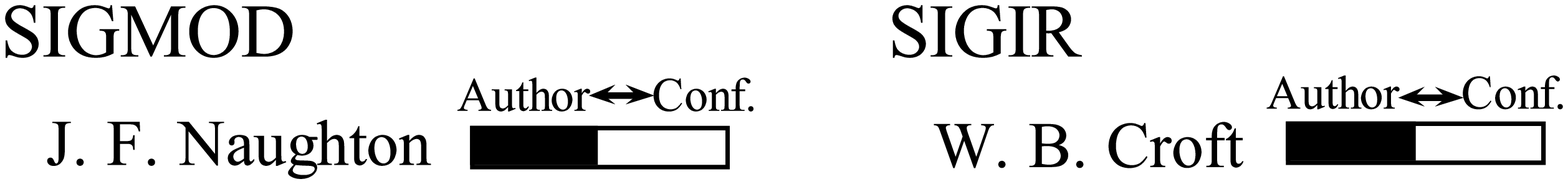}}\hspace{20pt}
    \subfigure[Asymmetric measure]
    {\includegraphics[scale=0.25]{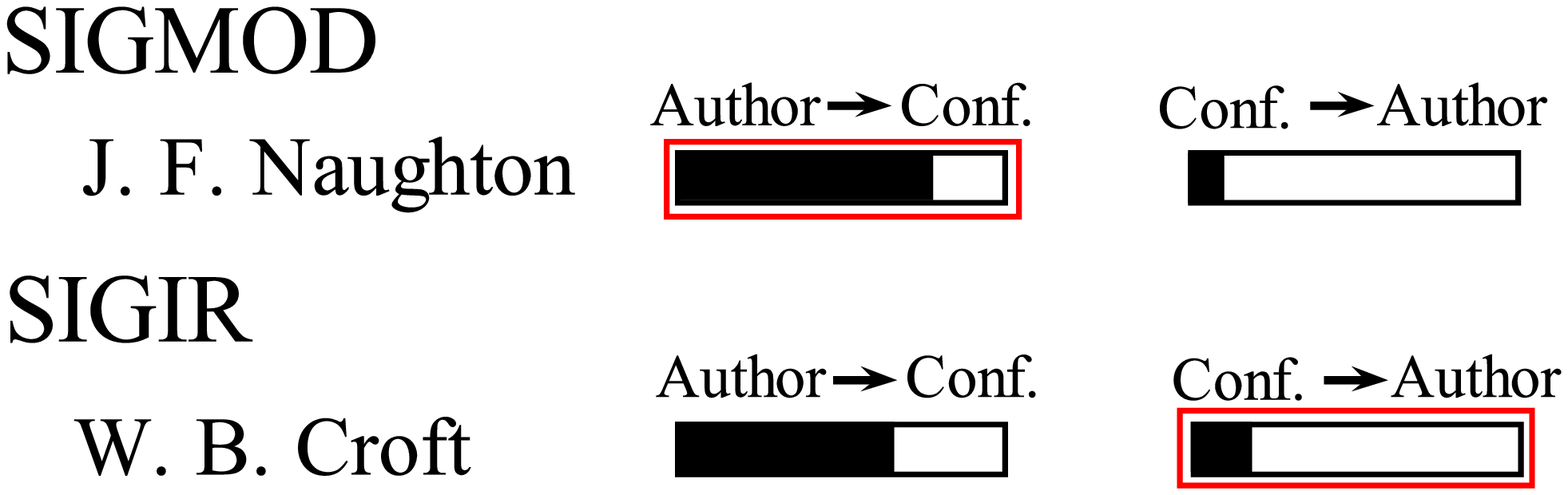}}
    \end{center}
    \caption{Examples of relative importance representing by symmetric and asymmetric measures. The rectangle with partially marked black denotes the
    relatedness of two objects.}\label{SymmRel}
\end{figure}

In this paper, we study the relevance search problem in
heterogeneous information networks. The aim of relevance search is
to effectively measure the relatedness of heterogeneous objects
(including objects with the same type or different types). Different
from the similarity search which only measures the similarity of
objects with same type, the relevance search measures the relatedness of
heterogeneous objects, not limit to objects with same type. Distinct from relational retrieval \cite{ZVDI08, LC10} in information retrieval domain, here relevance search is done on heterogeneous networks which can be constructed from meta-data of objects. Moreover, we think that a desirable relevance measure should satisfy the symmetry property based on the following reasons. (1) The symmetric measure is more general and useful in many learning tasks. Although the symmetry property is not necessary in the query task, it is essential for many important tasks, such as clustering and collaborative filtering. Moreover, it is the necessary condition for a metric. (2) The symmetric measure makes more sense in many applications, especially for the relatedness of heterogeneous object pairs. For example, in some applications, we need to answer the question like who has the similar importance to the conference SIGIR as J. F. Naughton to the
SIGMOD. Through comparing the relatedness of object pairs, we can
deduce the information of their relative importance. However, it
only can be done by the symmetric measure, not the asymmetric
measure. It can be explained by the example shown in Fig.
\ref{SymmRel}. For the symmetric measure, we can deduce that W. B.
Croft\footnote{http://ciir.cs.umass.edu/personnel/croft.html} has the same importance to SIGIR as J. F. Naughton\footnote{http://pages.cs.wisc.edu/$\sim$naughton/} to the
SIGMOD, since their relatedness scores are close. Suppose we know J.
F. Naughton is an influential researcher in SIGMOD, we can conclude that W. B.
Croft is also an influential researcher in SIGIR. However, we cannot deduce
the relative importance information from an asymmetric measure as
shown in Fig. \ref{SymmRel}(b). From the relatedness of author to
conference and conference to author, we will draw conflicting
conclusions.

Despite its value and significance, the relevance search in
heterogeneous networks has seldom been studied so far. It faces the
following research challenges. (1) Heterogeneous network is much more
complex than traditional homogeneous network. In heterogeneous
networks, different-typed objects and links are coexist in a network and they carry different semantic
meanings. As an bibliographic example shown in Fig. \ref{HIN}(b) (more details in Section V.A), it includes author, paper, term, and conference type. The relation ``author-paper'' means author writing paper, while the relation ``paper-conference'' means paper published in conference. If disregarding the difference of types and semantics, it does not make sense to mix different-typed objects to measure the similarity. We can find that search paths, connecting two objects through a sequence of relations between object types, embody rich semantic information \cite{SHYYW11}. Based on different search paths, the relatedness of two objects may be totally different. For
example, the relatedness of authors and conferences should be
different based on the ``author-paper-conference'' path and ``author-paper-author-paper-conference'' path, which mean the relations of authors publishing papers in
conferences and their co-authors publishing papers in
conferences, respectively. As a consequence, a desirable relevance measure should
be path-dependent, since such a measure can capture the semantics under paths and
return meaningful values based on different paths. (2) It is
difficult to design a uniform and symmetric relevance measure for
heterogeneous objects. In heterogeneous networks, the paths
connecting objects with same type are usually symmetric and the path length is a even number,
so it may be not difficult to design a symmetric measure based on the
symmetric paths, as the PathSim \cite{SHYYW11} does. However, the paths connecting objects with different types are asymmetric and the path length may be an odd number. In this condition, it is not easy to design a
symmetric relevance measure. It is more challengeable to design a uniform relevance measure for these two conditions.

Inspired by the intuition that two objects are related if they are
referenced by related objects, we propose a general framework, called
HeteSim, to evaluate the relatedness of heterogeneous objects in
heterogeneous networks. HeteSim is a path-based relevance measure, which can effectively capture the subtle semantics of search paths. Based on pair-wise random walk model, HeteSim treats arbitrary
search paths in a uniform way, which guarantees the symmetric
property of HeteSim. An additional benefit is that HeteSim can
evaluate the relatedness of objects with same or different types in the same way. Moreover, HeteSim is a semi-metric measure. In other words, HeteSim satisfies the properties of non-negativity,
identity of indiscernibles, and symmetry. It implies that HeteSim
can be used in many learning tasks (e.g., clustering and
collaborative filtering). We also consider the computation issue of HeteSim and propose four fast computation strategies. The extensive experiments validate the
effectiveness of HeteSim. As a general relevance measure, HeteSim illustrates its benefits and generality in knowledge discovery of heterogeneous networks through four case studies: automatically extracting object profile, experts finding through relative importance of object pairs, relevance search based on path semantics, and semantic-based movie recommendation. HeteSim also shows its potential in the machine learning tasks (i.e., query and clustering) where HeteSim outperforms other well-established similarity measures. In addition, numerous experiments test the significance of fast computing strategies of HeteSim.


\section{Related work}
The most related work to relevance search is similarity search. Here
we briefly summarize these works. Similarity search has been well
studied for a long time. These studies can be roughly categorized
into two types: feature based approaches and link based approaches.
The feature based approaches measure the similarity of objects based
on their feature values, such as cosine similarity, Jaccard
coefficient and Euclidean distance. The $k$ nearest neighbor is also
widely used in similarity measure \cite{BEKKS98,KS04}, which aims at
finding top-$k$ nearest neighbors according to similarities defined
on numerical features. Based on feature similarity, the top-$k$
similarity pair search algorithm (i.e., top-$k$-join) considers
similarity between tuples \cite{XWLS09}. This type of approaches
does not consider link relation among objects, so they cannot be
applied to networked data.

The link based approaches measure the similarity of objects based on
their link structures in a graph. The asymmetrical similarity
measure, Personalized PageRank \cite{JW03}, evaluates the
probability starting from a source object to a target object by
randomly walking with restart. It is extended to
the scalable calculation for online queries \cite{FRCS05,TFP06} and
the top-$k$ answers \cite{GPC08}. SimRank \cite{JW02} is a symmetric
similarity measure, which evaluates the similarity of two objects by
their neighbor's similarities. Because of its computational
complexity, many follow-up studies are done to accelerate such
calculations \cite{LHHJSYW10,LVGT08}. SCAN \cite{XYFTS07} measures
similarity of two objects by comparing their immediate neighbor
sets. Recently, Jin et al. proposed RoleSim to measure the role
similarity of node pair by automorphic equivalence \cite{JLH11}. These
approaches just consider the objects with the same type, so they can
not be applied in heterogeneous networks. ObjectRank \cite{BHP04}
applies authority-based ranking to keyword search in labeled graphs
and PopRank \cite{NZWM05} proposes a domain-independent object-level
link analysis model. Although these two approaches noticed that
heterogeneous relationships could affect the similarity, they do not
consider the distinct semantics of paths that include
different-typed objects, so they also cannot measure the similarity
of objects in heterogeneous networks.

Recently, the relevance research in heterogeneous
data emerge. Wang et al. \cite{WRFZHB11} proposed a model to
learn relevance from heterogeneous data, while their model more
focuses on analyzing the context of heterogeneous networks, rather
than network structure. Based on a Markov-chain model of random walk, Fouss et al. \cite{FPRS07} designed a similarity metric ECTD with nice properties and interpretation. Unfortunately, absent of path constraint, ECTD cannot capture the subtle semantics in heterogeneous networks. Considering semantics in meta paths
constituted by different-typed objects, Sun et al. \cite{SHYYW11}
proposed PathSim to measure the similarity of same-typed objects
based on symmetric paths. However, many valuable paths are
asymmetric and the relatedness of different-typed objects are also
meaningful. PathSim is not suitable in these conditions. In
information retrieval community, Lao and Cohen \cite{LC10,LC10a} proposed
a Path Constrained Random Walk (PCRW) model to measure the entity
proximity in a labeled directed graph constructed by the rich
metadata of scientific literature. Although the PCRW model can be
applied to measuring the relatedness of different-typed objects, the
asymmetric property of PCRW restricts its applications. In our HeteSim definition, users can measure the relatedness of heterogeneous objects based on an arbitrary search
path. The good merits of HeteSim (e.g., symmetric and self-maximum)
make it suitable for more applications.

\section{preliminary}

A heterogeneous information network is a special type of information
network, which either contains multiple types of objects or multiple types of
links.
\begin{defn}
\small \textbf{Information Network}. Given a schema
$S=(\mathcal{A},\mathcal{R})$ which consists of a set of entities
types $\mathcal{A}=\{A\}$ and a set of relations
$\mathcal{R}=\{R\}$, an information network is defined as a directed
graph $G=(V,E)$ with an object type mapping function $\phi:
V\rightarrow \mathcal{A}$ and a link type mapping function $\psi:
E\rightarrow \mathcal{R}$. Each object $v\in V$ belongs to one
particular object type $\phi(v)\in \mathcal{A}$, and each link $e
\in E$ belongs to a particular relation $\psi(e)\in \mathcal{R}$.
When the types of objects $|\mathcal{A}|>1$ or the types of
relations $|\mathcal{R}|>1$, the network is called
\textbf{heterogeneous information network}; otherwise, it is a
\textbf{homogeneous information network}.
\end{defn}

In information networks, we distinguish object types and relation
types. As a template for a network, the network schema depicts the
object types and the relations existing among object types. For a
relation $R$ existing from type $A$ to type $B$, denoted as
$A\overset{R}{\longrightarrow}B$, $A$ and $B$ are the \textbf{source
type} and \textbf{target type} of relation $R$, which is denoted as
$R.S$ and $R.T$, respectively. The inverse relation $\emph{R}^{-1}$
holds naturally for $B\overset{\emph{R}^{-1}}{\longrightarrow}A$.
Generally, $\emph{R}$ is not equal to $\emph{R}^{-1}$, unless
$\emph{R}$ is symmetric and these two types are the same.

\begin{figure}[t]
    \begin{center}
    \subfigure[ACM data]
    {\includegraphics[scale=0.2]{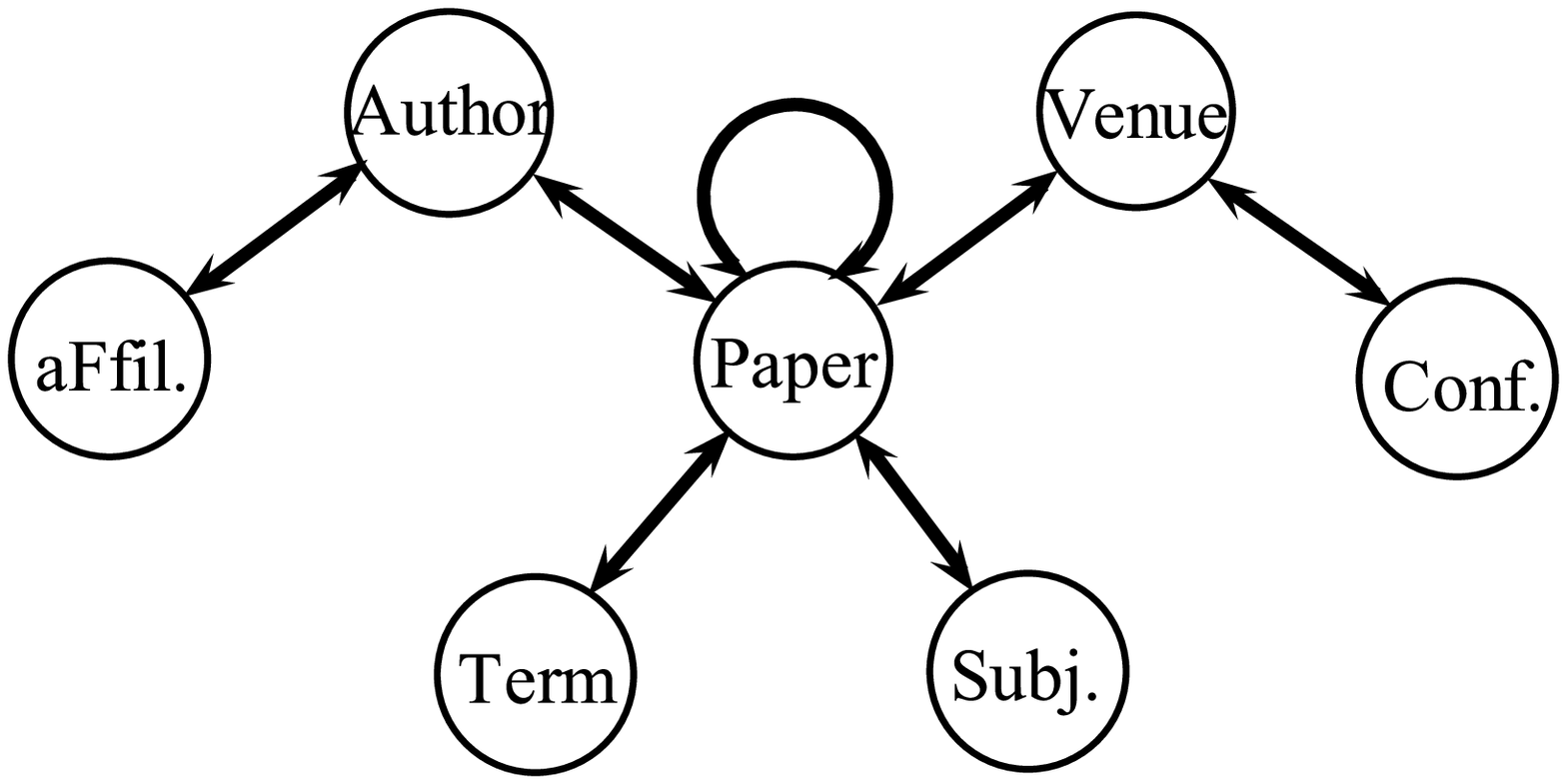}}\hspace{20pt}
    \subfigure[DBLP data]
    {\includegraphics[scale=0.22]{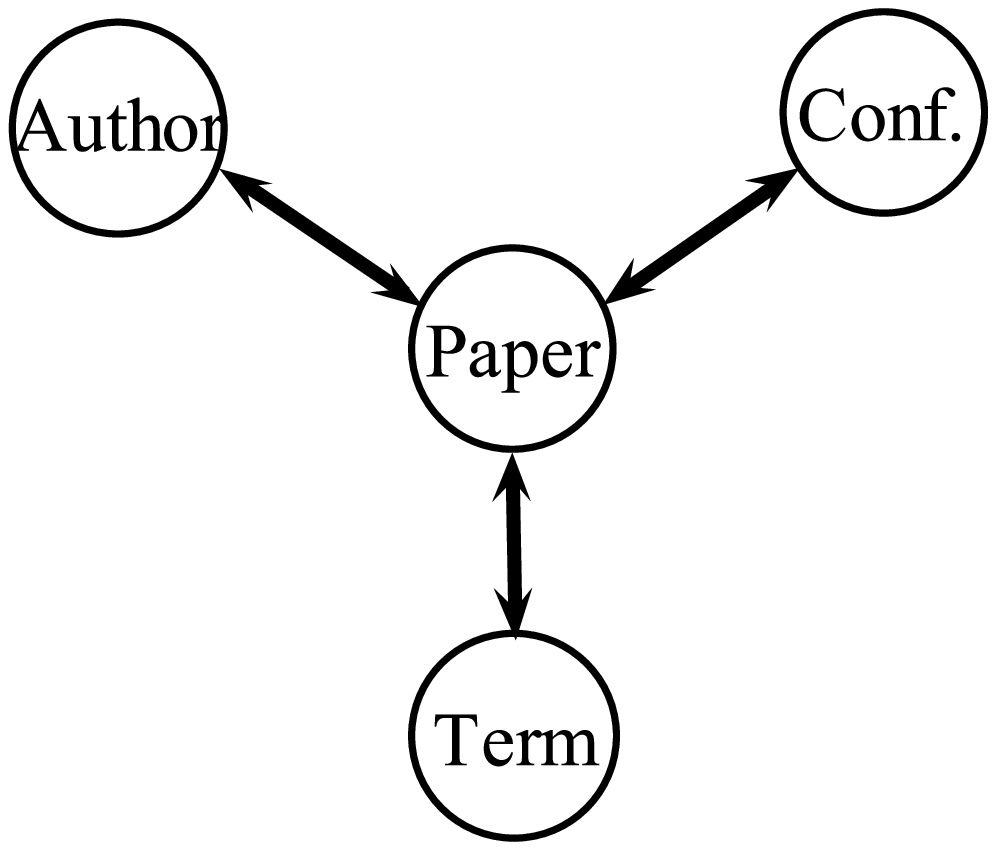}}\hspace{20pt}
    \subfigure[Movie data]
    {\includegraphics[scale=0.20]{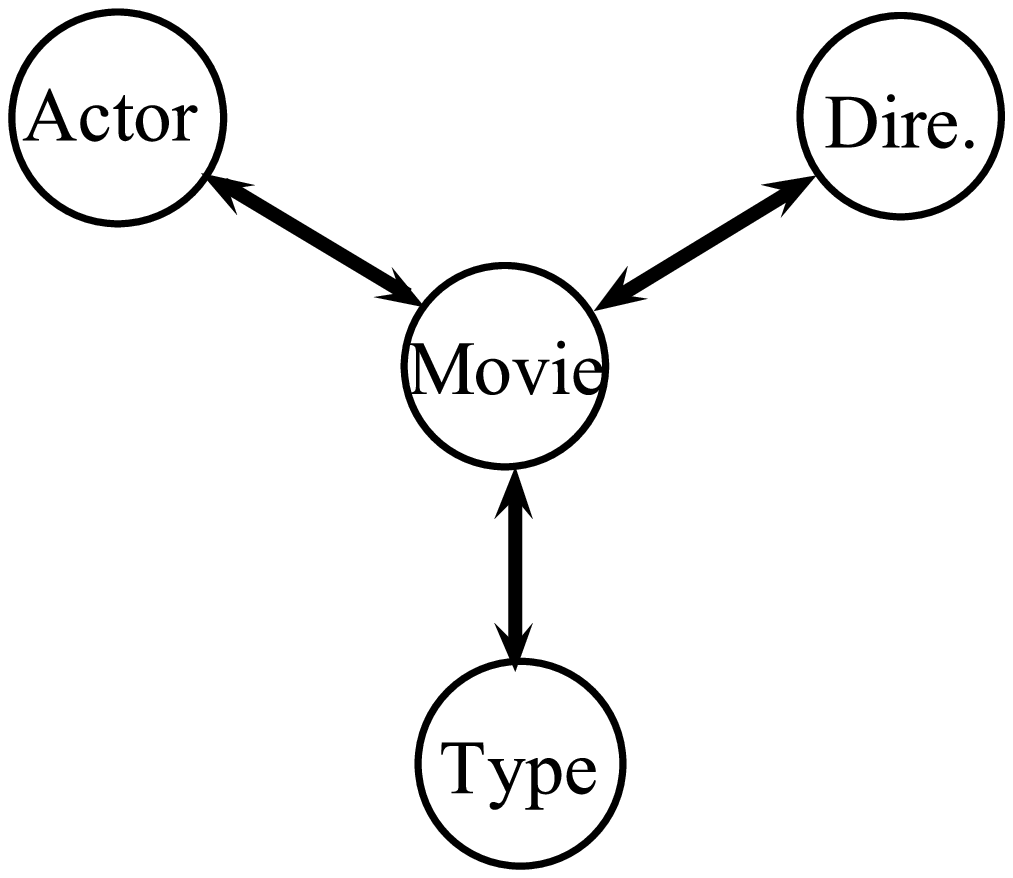}}
    \end{center}
    \caption{Examples of heterogeneous information network schema.}\label{HIN}
\end{figure}

\begin{exam}
\small A bibliographic information network is a typical heterogeneous
information network. The network schema of ACM dataset (see Section
V.A) is shown in Fig.\ref{HIN}(a). It contains objects from seven
types of entities: papers (P), authors (A), affiliations (F), terms
(T), subjects (S), venues (V), and conferences (C) (a conference
includes multiple venues, e.g., KDD including KDD2010, KDD2009 and
so on). There are links connecting different-typed objects. The link
types are defined by the relations between two object types. For
example, links exist between authors and papers denoting the writing
or written-by relations, between venues and papers denoting the
publishing or published-in relations. Fig.\ref{HIN}(b) and (c) show the
network schema of DBLP dataset and IMDB movie data (see Section V.A), respectively.
\end{exam}

Different from homogeneous networks, two objects in a heterogeneous
network can be connected via different paths and these paths have
different meanings. For example, in Fig. \ref{HIN}(a), authors and conferences can be connected via ``Author-Paper-Venue-Conference" (\emph{APVC}) path, ``Author-Paper-Subject-Paper-Venue-Conference"
(\emph{APSPVC}) path, and so on. The semantics
underneath these two paths are different. The \emph{APVC} path means
that papers written by authors are published in conferences, while
the \emph{APSPVC} path means that papers having the same subjects as
the authors' papers are published in conferences. Obviously, the
distinct semantics under different paths will lead to different
results. The relatedness under \emph{APVC} path emphasizes
the conferences that authors participated, while the relatedness
under \emph{APSPVC} path emphasizes on conferences publishing the
papers that have the same subjects with authors' papers. For
example, most of Christos Faloutsos's papers are published in the KDD,
VLDB, and SIGMOD. However, the papers having the same subjects with
his papers may be published in widespread conferences, such
as ICDM, SDM, and CIKM. So the relatedness of objects depends on the
search path in the heterogeneous networks. Formally, we define the
meta search path as the relevance path.

\begin{defn}
\small \textbf{Relevance Path}. A relevance path $\mathcal{P}$ is a
path defined on a schema $S=(\mathcal{A},\mathcal{R})$, and is
denoted in the form of $A_1
\overset{R_1}{\longrightarrow}A_2\overset{R_2}{\longrightarrow}\cdots
\overset{R_l}{\longrightarrow}A_{l+1}$ which defines a composite
relation $R=R_1\circ R_2\circ \cdots \circ R_{l}$ between type $A_1$
and $A_{l+1}$, where $\circ$ denotes the composition operator on
relations. The length of the path $\mathcal{P}$ is the number of
relations in $\mathcal{P}$, which is $l$.
\end{defn}

For simplicity, we can also use type names denoting the relevance
path if there are no multiple relations between the same pair of
types: $\mathcal{P}=(A_1A_2\cdots A_{l+1})$. We say a concrete path $p=(a_1a_2\cdots a_{l+1})$
between $a_1$ and $a_{l+1}$ in network $G$ is a \textbf{path
instance} of the relevance path $\mathcal{P}$, if for each $a_i$,
$\phi(a_i)=A_i$ and each link $e_i=\langle a_i, a_{i+1}\rangle$
belongs to the relation $R_i$ in $\mathcal{P}$. It can be denoted as
$p\in \mathcal{P}$. A relevance path $\mathcal{P}^{-1}$ is the
\textbf{reverse path} of $\mathcal{P}$, which defines an inverse
relation of the one defined by $\mathcal{P}$. Similarly, we define
the \textbf{reverse path instance} of $p^{-1}$ as the reverse path
of $p$ in $G$. Further, a relevance path
$\mathcal{P}$ is a \textbf{symmetric path}, if the relation $R$
defined by it is symmetric (i.e., $\mathcal{P}$ is equal to
$\mathcal{P}^{-1}$), such as $APA$ and $APCPA$. Two relevance paths
$\mathcal{P}_1=(A_1A_2\cdots A_l)$ and $\mathcal{P}_2=(B_1B_2\cdots
B_k)$ are \textbf{concatenable} if and only if $A_l$ is equal to $B_1$, and the
concatenated path is written as $\mathcal{P}=(\mathcal{P}_1\mathcal{P}_2)$, which equals to
$(A_1A_2\cdots A_lB_2\cdots B_k)$. A simple concatenable example is
that $AP$ and $PV$ can be concatenated to the path $APV$.

\section{HeteSim: A Uniform and Symmetric Relevance Measure}

\subsection{Basic Idea}
In many domains, similar objects are more likely to be related to some other similar objects. For
example, similar researchers usually publish many similar papers; similar
customers purchase similar commodities. As a consequence, two
objects are similar if they are referenced by similar objects. This
intuition is also fit for heterogeneous objects. For example,
 a researcher is more relevant to the conferences that the researcher has published papers in; and a customer is more faithful to the brands that the customer usually purchased.
Although the similar idea has been applied in
SimRank \cite{JW02}, it is limited to homogeneous networks. When we
apply the idea to heterogeneous networks, it faces the following
challenges. (1) The relatedness of heterogeneous objects is
path-constrained. The relevance path not only captures the semantics information but also constrains the walk path. So we need to design a path-based similarity measure. (2) A uniform and symmetric measure should be designed for arbitrary paths. For a given path (symmetric or asymmetric), the measure can evaluate the relatedness of heterogeneous object pair (same or different types) with one single score. In the following section, we will illustrate these challenges and their solutions in detail.


\subsection{Path-based Relevance Measure}
Different from homogeneous networks, the paths in heterogeneous
networks have semantics, which makes the relatedness of object pair depend on the given relevance path. Following the basic idea that similar objects are related to similar objects, we propose a path-based relevance measure: HeteSim.

\begin{defn}
\small\textbf{HeteSim}: Given a relevance path $\mathcal{P}=R_1\circ
R_2\circ \cdots \circ R_{l}$, the HeteSim score between two objects $s$ and
$t$ ($s\in R_1.S$ and $t\in R_l.T$) is:
\begin{equation}\label{sim}
 HeteSim(s,t|R_1\circ R_2\circ \cdots \circ
 R_{l})=\frac{1}{|O(s|R_1)||I(t|R_l)|}\sum_{i=1}^{|O(s|R_1)|}\sum_{j=1}^{|I(t|R_l)|}HeteSim(O_i(s|R_1),
 I_j(t|R_l)|R_2\circ  \cdots \circ R_{l-1})\\
\end{equation} 
where $O(s|R_1)$ is the out-neighbors of $s$ based on relation
$R_1$, and $I(t|R_l)$ is the in-neighbors of $t$ based on relation
$R_l$.
\end{defn}

When $s$ does not have any out-neighbors (i.e., $O(s|R_1)=\emptyset$)
or $t$ does not have any in-neighbors (i.e., $I(t|R_l)=\emptyset$)
following the path, we have no way to infer any relatedness between
$s$ and $t$ in this case, so we define their relevance value to be
0. Particularly, we consider objects with same type to have \textbf{self-relation} (denoted as $I$ relation) and each object only has self-relation with itself. It is obvious that
an object is just similar to itself for $I$ relation. So its relevance measure can be
defined as follows:

\begin{defn}
\small\textbf{HeteSim based on self-relation}: HeteSim between two
same-typed objects $s$ and $t$ based on the self-relation $I$ is:
\begin{equation}\label{IRelation}
HeteSim(s,t|I)=\delta(s,t)
\end{equation}
where $\delta(s,t)=1$, if $s$ and $t$ are same, or else $\delta(s,t)=0$.
\end{defn}

Equation (1) shows that the computation of $HeteSim(s,t|\mathcal{P})$
needs to iterate over all pairs $(O_i(s|R_1),$ $I_j(t|R_l))$ of
$(s,t)$ along the path ($s$ along the path and $t$ against path),
and sum up the relatedness of these pairs. Then, we normalize it by
the total number of out-neighbors of $s$ and in-neighbors of $t$.
That is, the relatedness between $s$ and $t$ is the average
relatedness between the out-neighbors of $s$ and the in-neighbors of
$t$. The process continues until $s$ and $t$ will meet along the
path. Similar to SimRank \cite{JW02}, HeteSim is also based on pair wise
random walk, while it considers the path constraint. As we know,
SimRank measures how soon two random surfers are expected to meet at
the same node \cite{JW02}. By contrast, $HeteSim(s,t|\mathcal{P})$
measures how likely $s$ and $t$ will meet at the same node when $s$
follows along the path and $t$ goes against the path.


\subsection{Decomposition of Relevance Path}
However, the source object $s$ and the target object $t$ may not
meet along a given path $\mathcal{P}$.
For the similarity measure of same-typed objects, the relevance
paths are usually even-length, even symmetric, so the source object
and the target object will meet at the middle objects. However, for the relevance measure of different-typed
objects, the relevance paths are usually odd-length. In this
condition, the source and target objects will never meet at the same
objects. Taking the $APVC$ path as an example, authors along the
path and conferences against the path will never meet in the same
objects. So the original HeteSim is not suitable for odd-length
relevance paths. In order to solve this difficulty, a basic idea is
to transform odd-length paths into even-length paths, and thus the
source and target objects are always able to meet at the same
objects. As a consequence, an arbitrary path can be decomposed as
two equal-length paths.

When the length $l$ of a relevance path $\mathcal{P}=(A_1A_2\cdots
A_{l+1})$ is even, the source objects (along the path) and the
target objects (against the path) will meet in the \textbf{middle
type} object $M=A_{\frac{l}{2}+1}$ on the \textbf{middle position}
$mid=\frac{l}{2}+1$, so the relevance path $\mathcal{P}$ can be
divided into two equal-length path $\mathcal{P}_L$ and
$\mathcal{P}_R$. That is, $\mathcal{P}=\mathcal{P}_L\mathcal{P}_R$,
where $\mathcal{P}_L=A_1A_2\cdots A_{mid-1}M$ and
$\mathcal{P}_R=MA_{mid+1}\cdots A_{l+1}$.

When the path length $l$ is odd, the source objects and the target
objects will meet at the relation
$A_{\frac{l+1}{2}}A_{\frac{l+1}{2}+1}$. For example, based on the
\emph{APSPVC} path, the source and target objects will meet at the
$SP$ relation after two steps. In order to let the source and target
objects meet at same-typed objects, we can add a middle type object
\emph{E} between the atomic relation
$A_{\frac{l+1}{2}}A_{\frac{l+1}{2}+1}$ and maintain the relation
between $A_{\frac{l+1}{2}}$ and $A_{\frac{l+1}{2}+1}$ at the same
time. Then the new path becomes $\mathcal{P'}=(A_1\cdots E\cdots
A_{l+1})$ which length is $l+1$, an even number. In the
aforementioned example, the path becomes \emph{APSEPVC}, whose length is
even now. The source objects and the target objects will meet
in the \textbf{middle type} object $M=E$ on the \textbf{middle
position} $mid=\frac{l+1}{2}+1$. As a consequence, the new relevance
path $\mathcal{P'}$ can also be decomposed into two equal-length path
$\mathcal{P}_L$ and $\mathcal{P}_R$.

\begin{defn}
\small \textbf{Decomposition of relevance path}. An arbitrary
relevance path $\mathcal{P}=(A_1A_2\cdots A_{l+1})$ can be
decomposed into two equal-path path $\mathcal{P}_L$ and
$\mathcal{P}_R$ (i.e., $\mathcal{P}=\mathcal{P}_L\mathcal{P}_R$),
where $\mathcal{P}_L=A_1A_2\cdots A_{mid-1}M$ and
$\mathcal{P}_R=MA_{mid+1}\cdots A_{l+1}$. $M$ and $mid$ are defined
as above.
\end{defn}

Obviously, for a symmetric path
$\mathcal{P}=\mathcal{P}_L\mathcal{P}_R$, $\mathcal{P}^{-1}_R$ is
equal to $\mathcal{P}_L$. For example, the relevance path
$\mathcal{P}=APCPA$ can be decomposed as $\mathcal{P}_L=APC$ and
$\mathcal{P}_R=CPA$. For the relevance path \emph{APSPVC}, we can
add a middle type object $E$ in $SP$ and thus the path becomes
\emph{APSEPVC}, so $\mathcal{P}_L=APSE$ and $\mathcal{P}_R=EPVC$.

\begin{figure}[t]
    \begin{center}
    \subfigure[Add middle type object]
    {\includegraphics[scale=0.3]{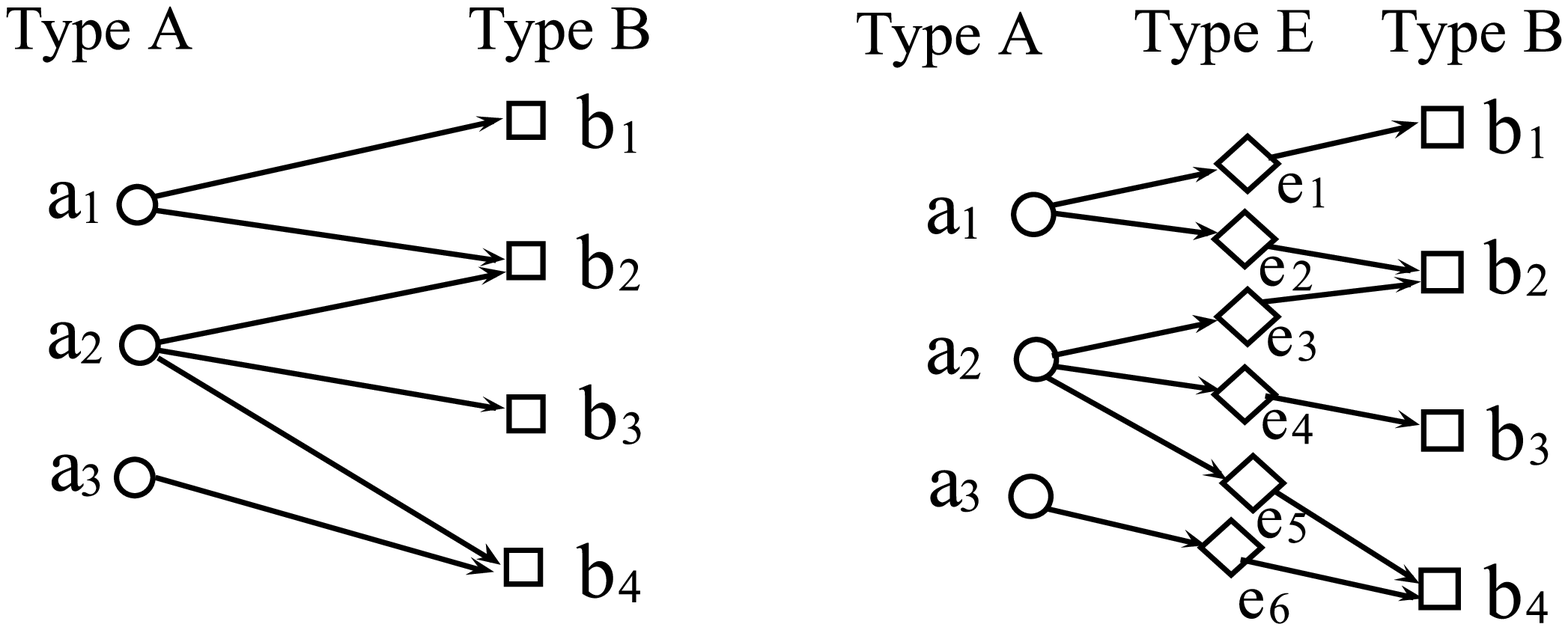}}
    \subfigure[Decomposition of atomic relation]
    {\includegraphics[scale=0.3]{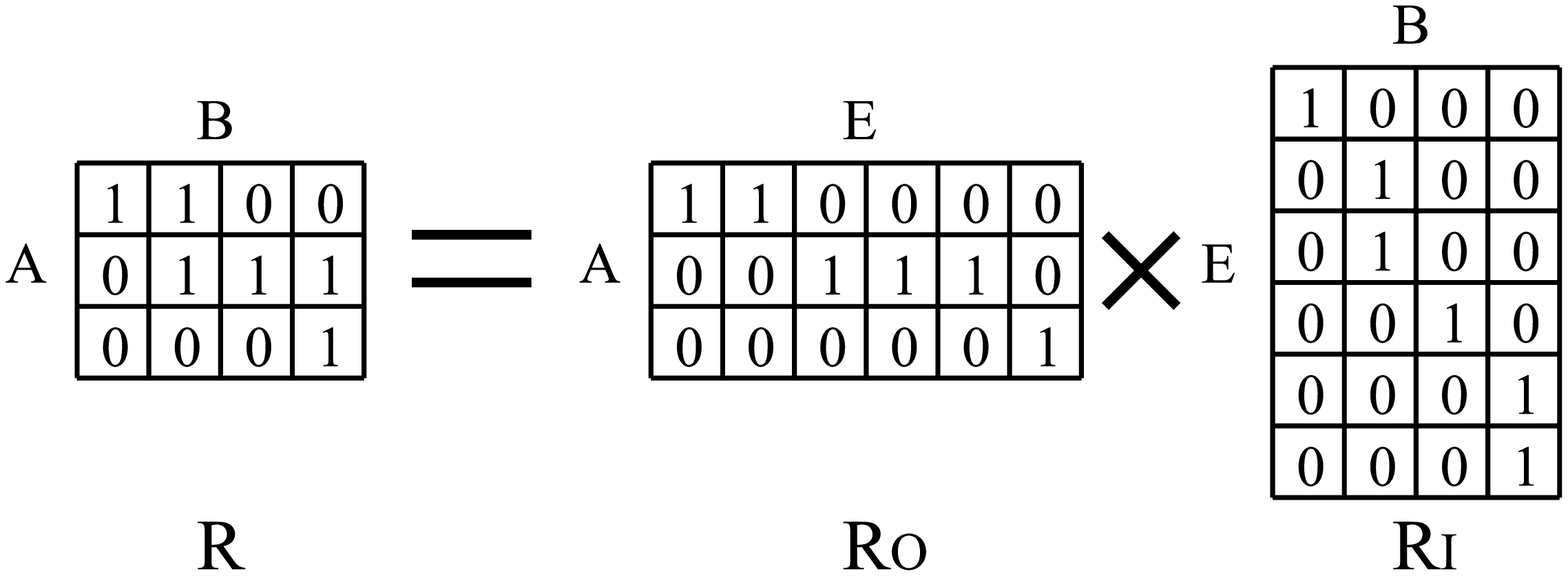}}
    \subfigure[HeteSim scores before normalization]
    {\includegraphics[scale=0.3]{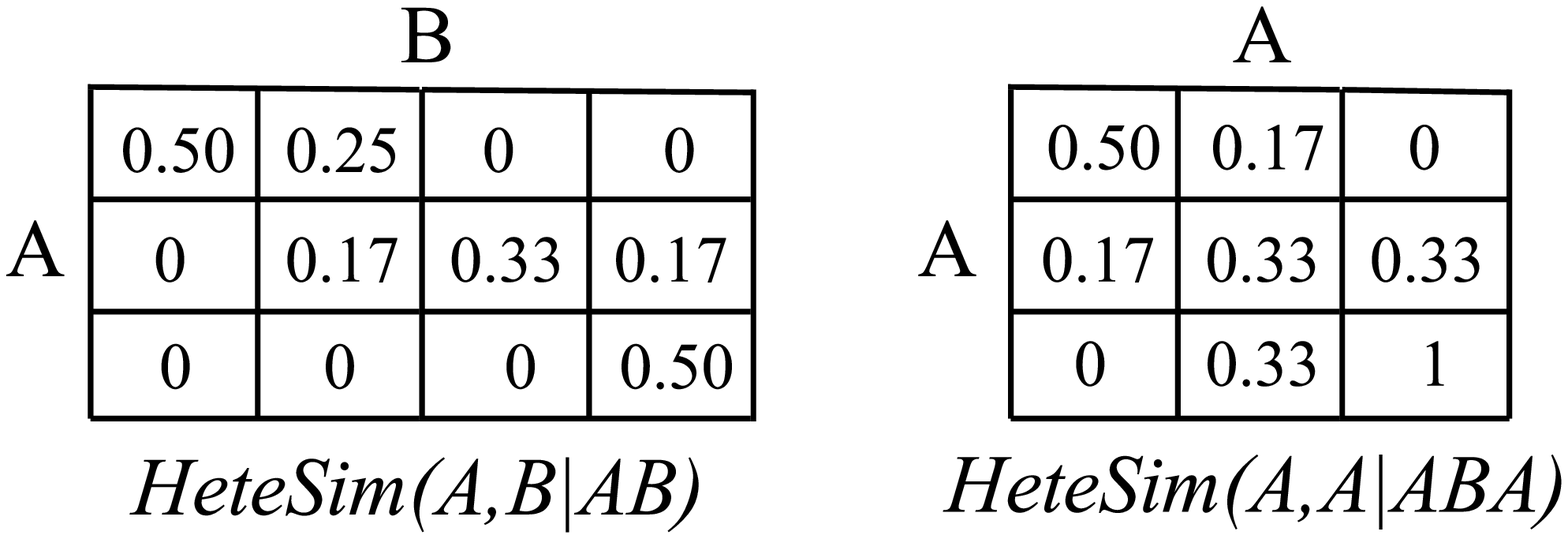}}
    \subfigure[HeteSim scores after normalization]
    {\includegraphics[scale=0.3]{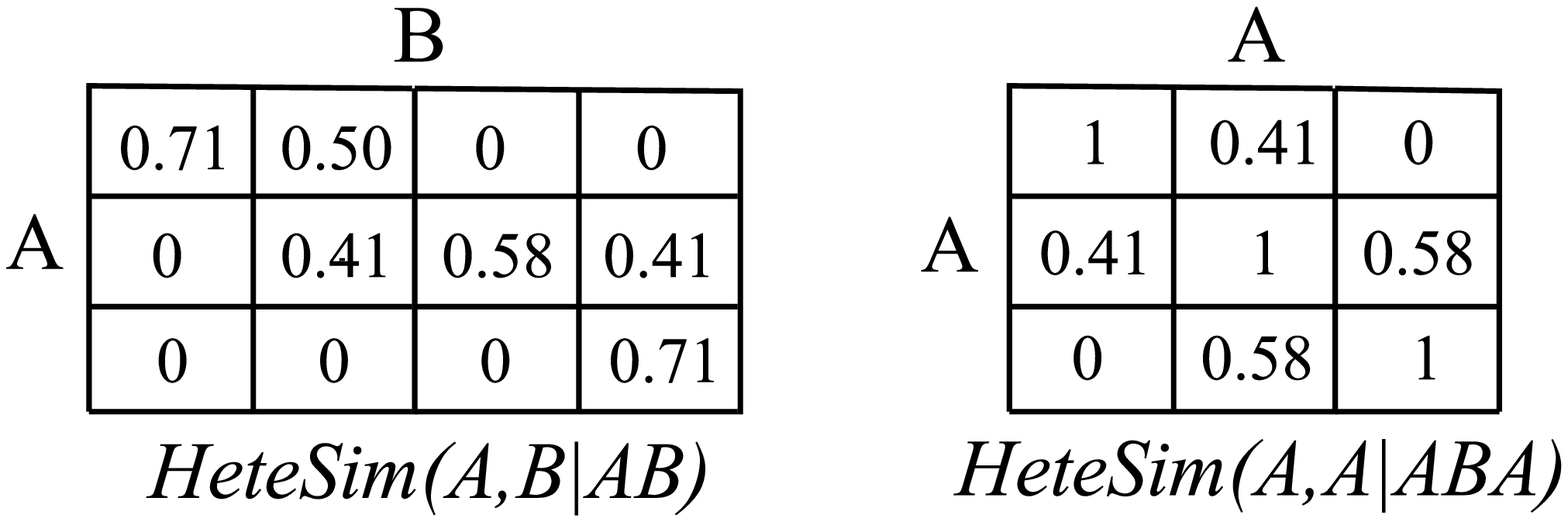}}
    \end{center}
    \caption{Decomposition of atomic relation and its HeteSim calculation.}\label{DecomExam}
\end{figure}

The next question is how we can add the middle type object \emph{E}
in an atomic relation \emph{R} between $A_{\frac{l+1}{2}}$ and
$A_{\frac{l+1}{2}+1}$ in an odd-length path. In order to contain
original atomic relation, we need to make the \emph{R} relation be
the composition of two new relations. To do so, for each instance of
relation \emph{R}, we can add an instance of \emph{E} to connect the
source and target objects of the relation instance. An example is
shown in Fig. \ref{DecomExam}(a), where the middle type object
\emph{E} is added in between the atomic relation \emph{AB} along
each path instance.

\begin{defn}
\small\textbf{Decomposition of atomic relation}. For an atomic relation
$R$, we can add an object type E (called edge object) between the
$R.S$ and $R.T$. And thus the atomic relation $R$ is decomposed as
$R_O$ and $R_I$ where $R_O$ represents the relation between $R.S$
and $E$ and $R_I$ represents that between $E$ and $R.T$. For each
relation instance $r\in R$, an instance $e\in E$ connects $r.S$ and
$r.T$. The paths $r.S\rightarrow e$ and $e\rightarrow r.T$ are the
instances of $R_O$ and $R_I$, respectively.
\end{defn}

It is clear that the decomposition has the following property, whose proof can be found in the Appendix A.

\textbf{Property 1.} An atomic relation $R$ can be decomposed as
$R_O$ and $R_I$, $R=R_O\circ R_I$, and this decomposition is unique.

Based on this decomposition, the relatedness of two objects with an
atomic relation $R$ can be calculated as follows:

\begin{defn}
\small \textbf{HeteSim based on atomic relation}: HeteSim between two
different-typed objects $s$ and $t$ based on an atomic relation $R$
($s\in R.S$ and $t\in R.T$) is:
\begin{equation}\label{HSbasedAR}
 HeteSim(s,t|R)=HeteSim(s,t|R_O\circ R_I)
 =\frac{1}{|O(s|R_O)||I(t|R_I)|}\sum_{i=1}^{|O(s|R_O)|}\sum_{j=1}^{|I(t|R_I)|}\delta(O_i(s|R_O),
 I_j(t|R_I))
\end{equation} 
\end{defn}

It is easy to find that $HeteSim(s,t|I)$ is a special case of
$HeteSim(s,t|R)$, since, for the self-relation $I$, $I=I_O\circ I_I$
and $|O(s|I_O)|=|I(t|I_I)|=1$. Definition 7 means that HeteSim can
measure the relatedness of two different-typed objects with an
atomic relation $R$ directly through calculating the average
of their mutual influence.

\begin{exam}
\small Fig. \ref{DecomExam}(a) shows an example of decomposition of atomic
relation. The relation $AB$ is decomposed into the relations $AE$
and $EB$. Moreover, the relation $AB$ is the composition of $AE$ and
$EB$ as shown in Fig. \ref{DecomExam}(b). Two HeteSim examples are illustrated in Fig. \ref{DecomExam}(c). We can find that HeteSim justly reflects relatedness of objects. Taking
$a_2$ for example, although $a_2$ equally connects with $b_2$,
$b_3$, and $b_4$, it is more close to $b_3$, because $b_3$ only
connects with $a_2$. This information is correctly reflected in the
HeteSim score of $a_2$ based on $AB$ path: $(0, 0.17, 0.33, 0.17)$.
\end{exam}

We also find that the similarity of an object and itself is not 1 in
HeteSim. Taking the right figure of Fig. \ref{DecomExam}(c) as example, the relatedness
of $a_2$ and itself is 0.33. It is obviously unreasonable. In the
following section, we will normalize the HeteSim and make the
relevance measure more reasonable.

\subsection{Normalization of HeteSim}
Firstly, we introduce the calculation of HeteSim between any two
objects given an arbitrary relevance path.

\begin{defn}
\small \textbf{Transition probability matrix}. For relation $A
\overset{R}{\longrightarrow}B$, $W_{AB}$ is an adjacent matrix
between type $A$ and $B$. $U_{AB}$ is a normalized matrix of $W_{AB}$
along the row vector, which is the transition probability matrix of
$A{\longrightarrow}B$ based on relation $R$. $V_{AB}$ is a normalized
matrix of $W_{AB}$ along the column vector, which is the transition
probability matrix of $B{\longrightarrow}A$ based on relation
$R^{-1}$.
\end{defn}

It is easy to prove that the transition probability matrix has the
following property. The proof can be found in the Appendix A.

\textbf{Property 2}. $U_{AB}=V'_{BA}$ and $V_{AB}=U'_{BA}$, where
$V'_{BA}$ is the transpose of $V_{BA}$.

\begin{defn}
\small \textbf{Reachable probability matrix}. Given a network
$G=(V,E)$ following a network schema $S=(\mathcal{A},\mathcal{R})$,
a reachable probability matrix $PM$ for a path
$\mathcal{P}=(A_1A_2\cdots A_{l+1})$ is defined as
$PM_{\mathcal{P}}=U_{A_1A_2}U_{A_2A_3}\cdots U_{A_{l}A_{l+1}}$ ($PM$
for simplicity). $PM(i,j)$ represents the probability of object
$i\in A_1$ reaching object $j\in A_{l+1}$ under the path
$\mathcal{P}$.
\end{defn}

According to the definition and Property 2 of HeteSim, the relevance between
objects in $A_1$ and $A_{l+1}$ based on the relevance path $\mathcal{P}=A_1A_2\cdots A_{l+1}$ is \\
\begin{equation}\small
\begin{split}
&HeteSim(A_1,A_{l+1}|\mathcal{P})=HeteSim(A_1,A_{l+1}|\mathcal{P}_L\mathcal{P}_R)\\
&=U_{A_1A_2}\cdots U_{A_{mid-1}M}V_{MA_{mid+1}}\cdots
V_{A_{l}A_{l+1}}\\
&=U_{A_1A_2}\cdots U_{A_{mid-1}M}U'_{A_{mid+1}M}\cdots U'_{A_{l+1}A_{l}}\\
&=U_{A_1A_2}\cdots U_{A_{mid-1}M}(U_{A_{l+1}A_{l}}\cdots U_{A_{mid+1}M})' \\
&=PM_{\mathcal{P}_L}PM'_{{\mathcal{P}_R}^{-1}}
\end{split}
\end{equation}

The above equation shows that the relevance of $A_1$ and $A_{l+1}$
based on the path $\mathcal{P}$ is the inner product of two probability
distributions that $A_1$ reaches the middle type object $M$ along
the path and $A_{l+1}$ reaches $M$ against the path. For two
instances $a$ and $b$ in $A_1$ and $A_{l+1}$, respectively, their
relevance based on path $\mathcal{P}$ is
\begin{equation}\label{}
HeteSim(a,b|\mathcal{P})=PM_{\mathcal{P}_L}(a,:)PM'_{{\mathcal{P}_R}^{-1}}(b,:)
\end{equation}
where $PM_{\mathcal{P}}(a,:)$ means the $a$-th row in
$PM_{\mathcal{P}}$.

We have stated that HeteSim needs to be normalized. It is reasonable
that the relatedness of the same objects is 1, so the HeteSim can be
normalized as follows:

\begin{defn}
\small \textbf{Normalization of HeteSim}. The normalized HeteSim
between two objects $a$ and $b$ based on the relevance path
$\mathcal{P}$ is:
\begin{equation}\label{}
HeteSim(a,b|\mathcal{P})=\frac{PM_{\mathcal{P}_L}(a,:)PM'_{{\mathcal{P}_R}^{-1}}(b,:)}{\sqrt{\|PM_{\mathcal{P}_L}(a,:)\|\|PM'_{{\mathcal{P}_R}^{-1}}(b,:)\|}}
\end{equation}

\end{defn}

In fact, the normalized HeteSim is the cosine of the probability
distributions of the source object $a$ and target object $b$
reaching the middle type object $M$. It ranges from 0 to 1. Fig.
\ref{DecomExam}(d) shows the normalized HeteSim scores. It is clear
that the normalized HeteSim is more reasonable. The normalization is an important step for HeteSim with the following advantages. (1) The normalized HeteSim has nice properties. The following Property 4 shows that  HeteSim satisfies the identity of indiscernibles. (2) It has nice interpretation. The normalized HeteSim is the cosine of two vectors representing reachable probability. As Fouss et al. pointed out \cite{FPRS07}, the angle between the node vectors is a much more predictive measure than the distance between the nodes.  In the following section, the HeteSim means the normalized HeteSim.

\subsection{Properties of HeteSim}

HeteSim has good properties, which makes it useful in many
applications. The proof of these properties can be found in the
Appendix A.

\textbf{Property 3}: \textbf{Symmetric}:
$HeteSim(a,b|\mathcal{P})=HeteSim(b,a|\mathcal{P}^{-1})$.

Property 3 shows the symmetric property of HeteSim. Although PathSim
\cite{SHYYW11} also has the similar symmetric property, it holds
only when the path is symmetric and $a$ and $b$ are with the same
type. The HeteSim has the more general symmetric property not only
for symmetric paths (note that $\mathcal{P}$ is equal to
$\mathcal{P}^{-1}$ for symmetric paths) but also for asymmetric
paths.

\textbf{Property 4}. \textbf{Self-maximum}:
$HeteSim(a,b|\mathcal{P})\in[0,1]$. $HeteSim(a,b|\mathcal{P})$ is
equal to 1 if and only if $PM_{\mathcal{P}_L}(a,:)$ is equal to
$PM_{{\mathcal{P}_R}^{-1}}(b,:)$.

Property 4 shows HeteSim is well constrained. For a symmetric path $\mathcal{P}$ (i.e., $\mathcal{P}_L={\mathcal{P}_R}^{-1}$), $PM_{\mathcal{P}_L}(a,:)$ is equal to
$PM_{{\mathcal{P}_R}^{-1}}(a,:)$, and thus $HeteSim(a,a|\mathcal{P})$ is equal to 1. If we define the distance
between two objects (i.e., $dis(s,t)$) as $dis(s,t)=1-HeteSim(s,t)$,
the distance of the same object is zero (i.e., $dis(s,s)=0$). As a
consequence, HeteSim satisfies the identity of indiscernibles. Note that it is a general identity of indiscernibles. For two objects with different types, their HeteSim score is also 1 if they have the same probability distribution on the middle type object. It is reasonable, since they have the similar structure based on the given path.

Since HeteSim obeys the properties of non-negativity, identity of
indiscernibles, and symmetry, we can say that HeteSim is a
semi-metric measure \cite{X09}. Because of a path-based
measure, HeteSim does not obey the triangle inequality. A
semi-metric measure has many good merits and can be widely used in many applications \cite{X09}.

\textbf{Property 5}. \textbf{Connection to SimRank}. For a bipartite
graph $G=(V,E)$ based on the
schema $S=(\{A,B\},\{R\})$, suppose the constant \emph{C} in SimRank is 1,\\
$SimRank(a_1,a_2)=\underset{n\rightsquigarrow
\infty}{lim}\sum_{k=1}^{n}{HeteSim(a_1,a_2|{(RR^{-1})}^{k})}$, \\
$SimRank(b_1,b_2)=\underset{n\rightsquigarrow
\infty}{lim}\sum_{k=1}^{n}{HeteSim(b_1,b_2|{(R^{-1}R)}^{k})}$. \\
where $a_1,a_2\in A$, $b_1,b_2\in B$ and
$A\overset{R}{\longrightarrow}B$. Here HeteSim is the non-normalized version.

This property reveals the connection of SimRank and HeteSim. SimRank
sums up the meeting probability of two objects after all possible
steps. HeteSim just calculates the meeting
probability along the given relevance path. If the relevance paths
explore all possible meta paths among the two objects, the sum of
HeteSim based on these paths is the SimRank. So we can say that HeteSim is a path-constrained version
of SimRank. Through relevance paths, HeteSim can subtly evaluate the similarity of heterogeneous objects with fine granularity. This property also implies
that HeteSim is more efficient than SimRank, since HeteSim only
needs to calculate the meeting probability along the given relevance
path, not all possible meta paths.

\subsection{Discussion}
Let us analyze the time and space complexity of computing HeteSim.
Suppose the average size of one type of objects is $n$ and there are
$T$ types objects, the space requirement of HeteSim is just $O(n^2)$
to store the relatedness matrix. Let $d$ be the average of
$|O(s|R_i)||I(t|R_j)|$ over all object-pairs $(s,t)$ based on
relation $R_i$ and $R_j$. For a given $l$-length relevance path, the
time required is $O(ldn^2)$, since node pairs (i.e., $n^2$)
calculate their relatedness along the relevance path. For SimRank,
the similarity of node pairs in all types (i.e., $(Tn)^2$) are
iteratively calculated at the same time, so its space complexity is
$O(T^2n^2)$, and the time complexity is $O(k(T^2d)(Tn)^2)$ (i.e.,
$O(kdn^2T^4)$), where $k$ is the number of iterations. So the
complexity of computing HeteSim is much smaller than SimRank.

Here, we discuss how to choose relevance path. There are several
ways to do it. (1) Users can select proper paths according to their
domain knowledge and experiences. (2) Supervised learning can be used to automatically determine the importance of relevance paths. In information retrieval field, Lao and Cohen \cite{LC10a} proposed a learnable proximity measure where proximity is defined by a weighted combination of simple ``path experts''. Through labeled training data, a learning algorithm can infer the weights of paths. The similar strategy can also be used for path selection. (3) Recently, Sun et al. \cite{SNHYYY12} combined meta path selection and user-guided information for clustering in heterogeneous networks. The similar user-guided information can also been applied in the selection of relevance paths in HeteSim.

There are numbers of similarity measures, most of which are based on three basic strategies \cite{SHYYW11}: (1) Path count strategy measures the number of path instances connecting source and target objects; (2) Random walk (RW) strategy measures the probability of the random walk from source to target objects; and (3) Pairwise random walk (PRW) strategy measures the pairwise random walk probability starting from source and target objects and reaching the same middle objects. Due to symmetry and arbitrary path constraints, we employ the PRW model in this work. Although the RW model can also satisfy the symmetric property through the combination of the reachable probability based on the paths $\mathcal{P}$ and $\mathcal{P}^{-1}$, it is redundancy for symmetric path, as well as short of nice interpretability. For the PRW model, it is inevitable to face the problem that the source and target object will not meet when the length of relevance path is odd. In order to solve it, some optional strategies can be applied, such as assigning the meeting object type. This paper adopts the path deposition strategy based on the following advantages. (1) It has a uniform framework to evaluate the relevance of same or different-typed objects for arbitrary paths. (2) It provides a simple but effective method to evaluate the relevance of two different-typed objects based on an atomic relation (see Def. 7).

Furtherly, we compare six well-established similarity measures in Table \ref{tab:COM}. There are three similarity measures for heterogeneous networks (i.e., HeteSim, PathSim, and PCWR) and three measures for homogeneous networks (i.e., P-PageRank, SimRank, and RoleSim), respectively. Although these similarity measures all evaluate the similarity of nodes by utilizing network structure, they have different properties and features. Three measures for heterogeneous networks all are path-based, since meta paths in heterogeneous networks embody semantics and simplify network structure. Two RW model based measures (i.e., P-PageRank and PCRW) do not satisfy the symmetric property. Because of satisfying the triangle inequation, RoleSim is a metric, while HeteSim, PathSim, and SimRank are semi-metric.

\begin{table*}
\centering \caption{Comparison of different similarity measures.}\label{tab:COM}
 \scriptsize
    \begin{tabular}{|c|c|c|c|c|c|}
        \hline
                   &Symmetry    &Triangle    &Path  &Model   &Features     \\
                   &            &Inequation  &based &        &                \\
        \hline
        HeteSim    &$\surd$  &$\times$   &$\surd$    &PRW      &evaluate relevance of heterogeneous objects based on arbitrary path\\
        \hline
        PathSim\cite{SHYYW11}    &$\surd$  &$\times$   &$\surd$    &Path Count & evaluate similarity of same-typed objects based on symmetric path      \\
        \hline
        PCWR\cite{LC10}       &$\times$  &$\times$   &$\surd$    &RW        &measure proximity to the query nodes based on given path    \\
        \hline
        SimRank\cite{JW02}   &$\surd$  &$\times$   &$\times$    &PRW        & measure similarity of node pairs based on the similarity of their neighbors \\
        \hline
        RoleSim\cite{JLH11}   &$\surd$  &$\surd$   &$\times$    &PRW         & measure real-valued role similarity based on automorphic equivalence \\
        \hline
        P-PageRank\cite{JW03}  &$\times$  &$\times$   &$\times$    &RW      & measure personalized views of importance based on linkage strcutre      \\
        \hline
    \end{tabular}
\normalsize
\end{table*}

\section{Experiments}

In the experiments, we validate the effectiveness of the HeteSim
on three datasets with four case studies and two learning tasks.

\subsection{Datasets}
Three heterogeneous information networks are employed in our experiments.

\textbf{ACM dataset:} The ACM dataset was downloaded from ACM
digital library\footnote{http://dl.acm.org/} in June 2010. The ACM
dataset comes from 14 representative computer science conferences:
KDD, SIGMOD, WWW, SIGIR, CIKM, SODA, STOC, SOSP, SPAA, SIGCOMM,
MobiCOMM, ICML, COLT, and VLDB. These conferences include 196
corresponding venue proceedings (e.g., KDD conference includes 12
proceedings, such as KDD'10, KDD'09, etc). The dataset has 12K
papers, 17K authors, and 1.8K author affiliations. After removing
stop words in the paper titles and abstracts, we get 1.5K terms that
appear in more than $1\%$ of the papers. The network also includes
73  subjects of these papers in ACM category. The network schema of
ACM dataset is shown in Fig. \ref{HIN}(a).

\textbf{DBLP dataset} \cite{JSDHG10}: The DBLP dataset is a
sub-network collected from DBLP
website\footnote{http://www.informatik.uni-trier.de/$\sim$ley/db/}
involving major conferences in four research areas: database, data
mining, information retrieval and artificial intelligence, which
naturally form four classes. The dataset contains 14K papers, 20
conferences, 14K authors and 8.9K terms, with a total number of 17K
links. In the dataset, 4057 authors, all 20 conferences and 100
papers are labeled with one of the four research areas. The network
schema is shown in Fig. \ref{HIN}(b).

\textbf{Movie dataset} \cite{SZKYLW12}: The IMDB movie data comes from
the Internet Movie Database \footnote{www.imdb.com/}, which includes movies, actors, directors and types. A movie heterogeneous network is constructed from the movie data and its schema
is shown in Fig. \ref{HIN}(c). The movie data contains 1.5K movies, 5K actors, 551 directors, and 112 types.

\subsection{Case Study}
In this section, we demonstrate the traits of HeteSim through case
study in four tasks: automatic object profiling, expert finding,
relevance search, and semantic recommendation.

\subsubsection{Task 1: Automatic Object Profiling}

We first study the effectiveness of our approach on different-typed
relevance measurement in the automatic object profiling task. If we
want to know the profile of an object, we can measure the relevance
of the object to objects that we are interested in. For example, we want to
know the academic profile of Christos
Faloutsos\footnote{http://www.cs.cmu.edu/$\sim$christos/}. It can be
solved through measuring the relatedness of Christos Faloutsos with
related objects, e.g., conferences, affiliations, other authors, etc.
Table \ref{tab:AutSearch} shows the lists of top relevant objects
with various types on ACM dataset. $APVC$ path shows the conferences
he actively participates. Note that KDD and SIGMOD are the two major
conferences Christos Faloutsos participates, which are mentioned in
his
homepage\footnote{http://www.cs.cmu.edu/$\sim$christos/misc.html}.
From the path \emph{APT}, we can obtain his research interests: data
mining, pattern discovery, scalable graph mining and social network.
Using $APS$ path, we can discover his research areas represented as
ACM subjects: database management (H.2) and data storage (E.2).
Based on $APA$ path, HeteSim finds the most important co-authors,
most of which are his Ph.D students. Another interesting case can be seen in Appendix B.
%
%
%

\begin{table*}
\centering \caption{Automatic object profiling task on author
``Christos Faloutsos" on ACM dataset.}\label{tab:AutSearch}
\scriptsize
    \begin{tabular}{|c|c|c|c|c|c|c|c|c|}
        \hline
         Path    &\multicolumn{2}{|c|}{\emph{APVC}}  & \multicolumn{2}{|c|}{\emph{APT}} &\multicolumn{2}{|c|}{\emph{APS}}&\multicolumn{2}{|c|}{\emph{APA}} \\
        \hline
        Rank & Conf.               & Score  & Terms  & Score  & Subjects       & Score  & Authors & Score  \\ \hline
        1    & KDD    & 0.1198 & mining       & 0.0930  & H.2 (database management)    & 0.1023 & Christos Faloutsos   & 1      \\
        2    & SIGMOD & 0.0284 & patterns     & 0.0926 & E.2 (data storage representations)   & 0.0232 & Hanghang Tong        & 0.4152\\
        3    & VLDB   & 0.0262 & scalable     & 0.0869 & G.3 (probability and statistics)  & 0.0175 & Agma Juci M. Traina  & 0.3250  \\
        4    & CIKM   & 0.0083 & graphs       & 0.0816 & H.3 (information storage and retrieval)   & 0.0136 & Spiros Papadimitriou & 0.2785 \\
        5    & WWW    & 0.0060  & social       & 0.0672 & H.1 (models and principles)  & 0.0135 & Caetano Traina, Jr.  & 0.2680  \\
        \hline
    \end{tabular}
\normalsize
\end{table*}

\begin{table*}
\centering \caption{Relatedness values of authors and conferences
measured by HeteSim and PCRW on ACM dataset.}\label{tab:RelImport}
\scriptsize
    \begin{tabular}{|rl|c|rl|c|rl|c|}
        \hline
        \multicolumn{3}{|c|}{HeteSim}&\multicolumn{6}{|c|}{PCRW}\\
        \hline
        \multicolumn{3}{|c|}{\emph{APVC}\&\emph{CVPA}}&\multicolumn{3}{|c|}{\emph{APVC}}&\multicolumn{3}{|c|}{\emph{CVPA}}\\
        \hline
        \multicolumn{2}{|c|}{Pair}   &Score   & \multicolumn{2}{|c|}{Pair}   &Score  & \multicolumn{2}{|c|}{Pair}   &Score  \\
        \hline
        C. Faloutsos,&KDD   &\textbf{0.1198}   & C. Faloutsos,&KDD &0.5517  & KDD,&C. Faloutsos&\textbf{0.0087}   \\
        \hline
        W. B. Croft,&SIGIR &\textbf{0.1201}  & W. B. Croft,&SIGIR &0.6481 & SIGIR,&W. B. Croft&\textbf{0.0098}\\
        \hline
        J. F. Naughton,&SIGMOD &\textbf{0.1185}   & J. F. Naughton,&SIGMOD &\textbf{0.7647}  & SIGMOD,&J. F. Naughton&0.0062   \\
         \hline
        A. Gupta,&SODA &\textbf{0.1225}    & A. Gupta,&SODA &\textbf{0.7647}  & SODA,&A. Gupta&\textbf{0.0090}   \\
        \hline
        Luo Si,&SIGIR &0.0734   & Luo Si,&SIGIR&\textbf{0.7059} & SIGIR,&Luo Si&0.0030    \\
         \hline
        Yan Chen,&SIGCOMM &0.0786  & Yan Chen,&SIGCOMM &\textbf{1}  & SIGCOMM,&Yan Chen&0.0013   \\
        \hline
    \end{tabular}
\normalsize
\end{table*}

\subsubsection{Task 2: Expert Finding}
In this case, we want to validate the effectiveness of HeteSim to
reflect the relative importance of object pairs through an expert
finding task. As we know, the relative importance of object pairs
can be revealed through comparing their relatedness. Suppose we know
the experts in one domain, the expert finding task here is to find
experts in other domains through their relative importances. Table
\ref{tab:RelImport} shows the relevance scores returned by different
approaches on six ``conference-author" pairs on ACM dataset. The
relatedness of conferences and authors are defined based on the
\emph{APVC} and \emph{CVPA} paths which have the same semantics:
authors publishing papers in conferences. Due to the symmetric
property, HeteSim returns the same value for both paths, while
PCRW returns different values for these two paths. Suppose that we
are familar with data mining area, and already know that C.
Faloutsos is an influential researcher in KDD. Comparing these
HeteSim scores, we can find influential researchers in other
research areas even if we are not quite familiar with these areas.
J. F. Naughton, W. B. Croft and A. Gupta should be influential
researchers in SIGMOD, SIGIR and SODA, respectively, since they have
very similar HeteSim score to C. Faloutsos. Moreover, we can also
deduce that Luo Si and Yan Chen may be active researchers in SIGIR
and SIGCOMM, respectively, since they have moderate HeteSim scores. In fact, C.
Faloutsos, J. F. Naughton, W. B. Croft and A. Gupta are top ranked
authors in their research communities. Luo Si and Yan Chen are the
young professors and they have done good work in their research
areas. However, if the relevance measure is not symmetric (e.g.,
PCRW), it is very hard to tell which authors are more influential
when comparing these relevance scores. For example, the PCRW score
of Yan Chen and SIGCOMM is the largest one in the \emph{APVC} path.
However, the value is the smallest one when the opposite path (i.e., \emph{CVPA} path) is
considered. A quantitative experiment in the Appendix C illustrates that, compared to PCRW, HeteSim can reveal the relative importance of author-conference
pairs more accurately.


\begin{table*}
\centering \caption{Top 10 related authors to ``Christos Faloutsos"
based on $APVCVPA$ path on ACM
dataset.}\label{tab:SameType}\scriptsize
        \begin{tabular}{|c|c|c|c|c|c|c|c|c|}
        \hline
             &\multicolumn{2}{|c|}{HeteSim} &\multicolumn{2}{|c|}{PathSim}  & \multicolumn{2}{|c|}{PCRW}  & \multicolumn{2}{|c|}{SimRank} \\
                \cline{2-9}
                Rank & Author                                     & Score   & Author                                                       & Score   & Author                         & Score  & Author                         & Score \\ \hline
                1           & Christos Faloutsos             & 1           & Christos Faloutsos                & 1           & Charu C. Aggarwal   & 0.0063  & Christos Faloutsos	&1\\
                2           & Srinivasan Parthasarathy        & 0.9937     & Philip Yu                         & 0.9376 & Jiawei Han                 & 0.0061 &Edoardo Airoldi	&0.0789 \\
                3           & Xifeng Yan                             & 0.9877 & Jiawei Han                        & 0.9346 & Christos Faloutsos & 0.0058 &Leejay Wu	&0.0767\\
                4           & Jian Pei                                 & 0.9857 & Jian Pei                          & 0.8956 & Philip Yu                   & 0.0056 &Kensuke Onuma	&0.0758\\
                5           & Jiong Yang                             & 0.9810   & Charu C. Aggarwal                 & 0.7102 & Alia I. Abdelmoty   & 0.0053 &Christopher R. Palmer	&0.0699\\
                6           & Ruoming Jin                           & 0.9758 & Jieping Ye                        & 0.6930   & Chris B. Jones         & 0.0053 &Anthony Brockwell	&0.0668\\
                7           & Wei Fan                                   & 0.9743 & Heikki Mannila                    & 0.6928 & Jian Pei                     & 0.0034 &Hanghang Tong &0.0658\\
                8           & Evimaria Terzi                     & 0.9695 & Eamonn Keogh                      & 0.6704 & Heikki Mannila         & 0.0032 &Evan Hoke &0.0651\\
                9           & Charu C. Aggarwal               & 0.9668 & Ravi Kumar                        & 0.6378 & Eamonn Keogh             & 0.0031 &Jia-Yu Pan	&0.0650\\
                10         & Mohammed J. Zaki                 & 0.9645 & Vipin Kumar                       & 0.6362 & Mohammed J. Zaki     & 0.0027 &Roberto Santos Filho	&0.0648\\
                \hline
        \end{tabular}
        \normalsize
\end{table*}

\subsubsection{Task 3: Relevance Search based on Path Semantics}
As we have stated, the path-based relevance measure can capture the
semantics of paths. In this relevance search task, we will observe the importance of paths and
the effectiveness of semantics capture through the comparison of three path-based measures (i.e., HeteSim, PCRW, and PathSim) and SimRank. This task
is to find the top 10 related authors to Christos Faloutsos based on
the $APVCVPA$ path which means authors publishing papers in same
conferences. Through ignoring the heterogeneity of objects, we directly run SimRank on whole network and select top ten authors from the rank results which mix different-typed objects together. The comparison results are shown in Table \ref{tab:SameType}. At first sight, we can find that three path-based measures  all return researchers having the similar reputation with Christos in slightly different orders. However, the results of SimRank are totally against our common sense. We think the reason of bad performances is that SimRank only considers link structure but ignores the link semantics. In heterogeneous networks, different-typed objects are connected together. If ignoring the link semantics and treating different-typed links equally, it will be full of noise. Through selecting useful relation sequences, the meta path avoids the noise caused by complex structure. Moreover, the meta path embody the semantics of relation sequence. As a consequence, the meta path is a basic analysis tool in heterogeneous networks.

\begin{figure}[t]
    \begin{center}
    {\includegraphics[scale=0.25]{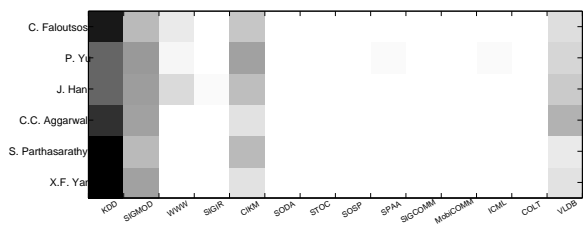}}
    \end{center}
    \caption{Probability distribution of authors' papers on 14 conferences of ACM dataset. }\label{ProbDist}
\end{figure}

In addition, let's analyze the subtle differences of results returned by three path-based measures. The PathSim finds the similar peer authors, such as Philip Yu and Jiawei
Han. They have the same reputation in data mining field. It is
strange for PCRW that the most similar author to Christos Faloutsos
is not himself, but Charu C. Aggarwal and Jiawei Han. It is
obviously not reasonable. Our conjecture is that Charu C. Aggarwal
and Jiawei Han published many papers in the conferences that
Christos Faloutsos participated in, so Christos Faloutsos has more
reachable probability on Charu C. Aggarwal and Jiawei Han than
himself along the $APVCVPA$ path. HeteSim's results are a little
different. The most similar authors are Srinivasan Parthasarathy and
Xifeng Yan, instead of Philip Yu and Jiawei Han.
Let's revisit the semantics of the path $APVCVPA$: authors
publishing papers in the same conferences. Fig. \ref{ProbDist} shows
the reachable probability distribution from authors to conferences
along the path $APVC$. It is clear that the probability distribution
of papers of Srinivasan Parthasarathy and Xifeng Yan on conferences
are more close to that of Christos Faloutsos, so they should be more
similar to Christos based on the same conference publication.
Although Philip Yu and Jiawei Han have the same reputation with C.
Faloutsos, their papers are more broadly published in different
conferences. So they are not the most similar authors to C.
Faloutsos based on the $APVCVPA$ path. As a consequence, our HeteSim
more accurately captures the semantics of the path. One more case in the Appendix D further illustrates the capability of HeteSim to capture the semantics of relevance paths.

\subsubsection{Task 4: Semantic Recommendation}
In this case study, we illustrate the potential of applying HeteSim in recommendation systems. An important goal of recommendation systems is to recommend products according to user's intent. The ideal recommendation system should be able to capture the subtlety of intents from different users. Take the movie dataset for example. Suppose that ``M'' represents movie, ``T'' represents the movie types. ``A'' and ``D" represent the actors and directors, respectively. If users want to find movies that share the same actors with ``Iron Man'', the \emph{MAM} path can be used in the recommendation system. For users who like the movies which are of the same type with ``Iron Man'', the path \emph{MTM} can be used. The recommendation results are illustrated in Table \ref{tab:MovieRec}. It is shown that HeteSim can recommend different movies based on different paths. The \emph{MAM} path recommends movies which share actors with the movie ``Iron Man'', such as ``The Kite Runner'' and ``The Good Night''. Although the first four recommended movies (except ``Iron Man'' itself) all have only one common actor with ``Iron Man'', the Kite Runner has less actors, so its score is higher. The \emph{MTM} path recommends movies of the same type with ``Iron Man'', such as ``The Incredible Hulk'', ``Teenage Mutant Turtles'', and ``Spawn''. ``The Incredible Hulk'' has more common types with ``Iron Man'', so it ranks the top one. More interestingly, based on relevance paths, the \emph{HeteSim} can recommend objects of different types. For example, a user may like the movies that have the same type with the movies of the actor ``Sylvester Stallone''. The \emph{AMTM} path can be adopted. The results are shown in the last column of Table \ref{tab:MovieRec}. Since ``Sylvester Stallone'' has acted the leading role in many movies about boxing and sport, the HeteSim recommends this kind of movies, such as ``Rocky'' and ``Million Dollar Baby''. Following this idea, we have designed a semantic-based recommendation system HeteRecom \cite{SZKYLW12}.

\begin{table*}
\centering \caption{Semantic recommendation on movie data.}\label{tab:MovieRec}
\scriptsize
\begin{tabular}{|l|l|l|l|l|l|l|}\hline
\multicolumn{1}{|c|}{} & \multicolumn{2}{c|}{ Search: Iron Man; Path: \emph{MAM} } & \multicolumn{2}{c|}{ Search: Iron Man; Path: \emph{MTM} } & \multicolumn{2}{c|}{ Search: Sylvester Stallone; Path: \emph{AMTM} } \\
\hline
\multicolumn{1}{|c|}{Rank} & \multicolumn{1}{c|}{Movie} & \multicolumn{1}{c|}{Score} & \multicolumn{1}{c|}{Movie} & \multicolumn{1}{c|}{Score} & \multicolumn{1}{c|}{Movie} & \multicolumn{1}{c|}{Score} \\
\hline
\multicolumn{1}{|c|}{1} & \multicolumn{1}{c|}{Iron Man} & \multicolumn{1}{c|}{1.0000} & \multicolumn{1}{c|}{Iron Man} & \multicolumn{1}{c|}{1.0000} & \multicolumn{1}{c|}{Rocky} & \multicolumn{1}{c|}{0.1023} \\
\hline
\multicolumn{1}{|c|}{2} & \multicolumn{1}{c|}{The Kite Runner} & \multicolumn{1}{c|}{0.2185} & \multicolumn{1}{c|}{The Incredibale Hulk} & \multicolumn{1}{c|}{0.8752} & \multicolumn{1}{c|}{Million Dollor Baby} & \multicolumn{1}{c|}{0.0981} \\
\hline
\multicolumn{1}{|c|}{3} & \multicolumn{1}{c|}{The Good Night} & \multicolumn{1}{c|}{0.1894} & \multicolumn{1}{c|}{TMNT} & \multicolumn{1}{c|}{0.8531} & \multicolumn{1}{c|}{The Wrestlet} & \multicolumn{1}{c|}{0.0932} \\
\hline
\multicolumn{1}{|c|}{4} & \multicolumn{1}{c|}{See Spot Run} & \multicolumn{1}{c|}{0.1894} & \multicolumn{1}{c|}{Spawn} & \multicolumn{1}{c|}{0.8256} & \multicolumn{1}{c|}{Hardball} & \multicolumn{1}{c|}{0.0895} \\
\hline
\multicolumn{1}{|c|}{5} & \multicolumn{1}{c|}{Proof} & \multicolumn{1}{c|}{0.1894} & \multicolumn{1}{c|}{Batman} & \multicolumn{1}{c|}{0.8171} & \multicolumn{1}{c|}{Out Cold} & \multicolumn{1}{c|}{0.0887} \\
\hline
\end{tabular}\normalsize
\end{table*}

%
%
%

\subsection{Performance on Query Task}
The query task will validate the effectiveness of HeteSim on query
search of heterogeneous objects. Since PathSim cannot measure the
relatedness of different-typed objects, we only
compare HeteSim with PCRW in this experiment. On DBLP dataset, we measure the proximity of conferences and authors based on the \emph{CPA} path. For each conference, we rank its
related authors according to their measure scores. Then we draw the ROC curve of top 100 authors according to the labels of authors (when the labels of author and conference are same, it is true, else it is false). After that, we calculate the AUC (Area Under ROC Curve) score to evaluate the performances of the ranked
results. Note that all conferences and some authors on the DBLP dataset are labeled with one of the four research areas (see Section V.A). The larger score means the better performance. We evaluate
the performances on 9 representative conferences and their AUC
scores are shown in Table \ref{tab:AUC}. We can find that HeteSim
consistently outperforms PCRW in all 9 conferences. It shows that
the proposed HeteSim method can work better
than the asymmetric similarity measure PCRW on proximity query task.

\begin{table*}
\centering \caption{AUC values for the relevance search of
conferences and authors based on \emph{CPA} path on DBLP
dataset.}\label{tab:AUC}
 \scriptsize
    \begin{tabular}{|c|c|c|c|c|c|c|c|c|c|c|c|c|c|c|c|c|c|c|c|}
        \hline
                  &KDD    &ICDM    &SDM     &SIGMOD     &VLDB      &ICDE  &AAAI   &IJCAI      &SIGIR          \\
        \hline
        HeteSim   &0.8111 &0.6752  &0.9504  &0.7662    &0.8262    &0.7322 &0.8110 &0.8754     &0.6132 \\

        \hline
        PCRW      &0.8030 &0.6731  &0.9390  &0.7588    &0.8200    &0.7263 &0.8067 &0.8712     &0.6068 \\
        \hline

    \end{tabular}
\normalsize
\end{table*}

\begin{table}
\centering \caption{Comparison of clustering performances for similarity
measures on DBLP dataset.}\label{tab:ClusAcc} \scriptsize
    \begin{tabular}{|c|c|c|c|c|c|c|c|c|}
        \hline
                  &\multicolumn{2}{|c|}{Venue NMI} &\multicolumn{2}{|c|}{Author NMI} &\multicolumn{2}{|c|}{Paper NMI} &Weighted &Running Time(s)\\
        \cline{2-7}
                  & Mean &Dev. & Mean &Dev. & Mean &Dev. &Avg. NMI  &on Author Clustering\\
        \hline
        HeteSim   &0.7683 &0.0716 &\textbf{0.7288} &0.0835 &\textbf{0.4989} &0.0675 &\textbf{0.7235} &7.5\\

        \hline
        PathSim   &0.8162 &0.1180 &0.6725 &0.1258 &0.3833 &0.1086 &0.6663 &50.7\\
        \hline
        PCRW      &0.7096	&0.0726	&0.7105	&0.0800	&0.4881	&0.0390 &0.7052 &2.3\\
        \hline
        SimRank	  &\textbf{0.8889}	&0.0928	&0.6854	&0.0662	&0.4694	&0.0319 &0.6812 &54\\
        \hline
        RoleSim   &0.2780	&0.0343	&0.5014	&0.0405	&0.3885	&0.0491 &0.4976 &55600\\
        \hline
        P-PageRank	&0.731	&0.0864	&0.4414	&0.001	&0.4212	&0.0637 &0.4447 &43\\
        \hline
    \end{tabular}
\normalsize
\end{table}

\subsection{Performance on Clustering Task}
Due to the symmetric property, HeteSim can be applied to clustering
tasks directly. In order to evaluate its performance, we compare HeteSim with five well-established similarity measures, including two path-based measures (i.e., PathSim and PCRW) and three homogeneous measures (i.e., SimRank, RoleSim, and P-PageRank). These measures use the same information to determine the pairwise similarity between objects. We evaluate the clustering performances
on DBLP dataset. There are three tasks: clustering on
conferences based on $CPAPC$ path, clustering on authors based on
$APCPA$ path, and clustering on papers based on $PAPCPAP$ path. For asymmetric measures (i.e., PCRW and P-PageRank), the symmetric similarity matrix can be obtained through the average of similarity matrix based on paths $\mathcal{P}$ and $\mathcal{P}^{-1}$. For RoleSim, it is applied in the network constructed by path $\mathcal{P}$. For SimRank and P-PageRank, they are applied in the subnetwork constructed by path $\mathcal{P}_{L}$ (note that three paths in experiments are symmetric). For example, for the $CPAPC$ path, the bipartite graph $M_{CA}$ derived from path $CPA$ can used in both SimRank and P-PageRank measures. Then we apply Normalized Cut \cite{SM00} to perform clustering based on the
similarity matrices returned by different measures. The number of
clusters is set as 4. The NMI criterion (Normalized Mutual Information)
\cite{SHZYCW09} is used to evaluate the clustering performances on
conferences, authors, and papers. NMI is between 0 and 1 and the
higher the better. In experiments, the damping factors for P-PageRank, SimRank, and RoleSim are set as 0.9, 0.8, and 0.1, respectively.

The average clustering accuracy results of 100 runs are summarized in Table \ref{tab:ClusAcc}. We can find that HeteSim achieves best performances on two tasks (authors and papers clustering) and third place on the conferences clustering task. In all, it performs best in terms of weighted average of clustering accuracy in three types. The mediocre results of PCWR and P-PageRank illustrate that, although symmetric similarity measures can be constructed by the combination of two random walk processes, the simple combination cannot generate good similarity measures. RoleSim aims to detect role similarity, a little different from structure similarity, so it has bad performances in these clustering tasks. In addition, we also record the running time for similarity computation of all measures. Due to space limitation, we only show the representative running time on author clustering task in the last column of Table \ref{tab:ClusAcc}. We can find HeteSim and PCWR have the smallest running time, since they only need to compute matrix multiplication once along the path. The iterative computation in SimRank and P-PageRank make them longer running time. The neighbor matching process in RoleSim has high time complexity, which makes it very time-consuming. The experiments show that HeteSim not only does well on similarity measure of same-typed objects but also has the potential as the similarity measure in clustering with high efficiency.

\section{Quick Computation Strategies and Experiments}
HeteSim has a high computation demand for time and space. It
is not affordable for on-line query in large-scale information
networks. So a primary strategy is to compute relevance matrix
off-line and do on-line queries with these matrix. For
frequently-used relevance paths, the relatedness matrix
$HeteSim(A,B|\mathcal{P})$ can be materialized ahead of time. The
on-line query on $HeteSim(a,B|\mathcal{P})$ will be very fast, since
it only needs to locate the row and column in the matrix. However,
it also costs much time and space to materialize all frequently-used
paths. As a consequence, we propose four strategies to fast compute
the relevance matrix. Moreover, experiments validate the effectiveness of these strategies.

\begin{figure*}[t]
    \begin{center}
    \subfigure[Running time in MUL]
    {\includegraphics[scale=0.3]{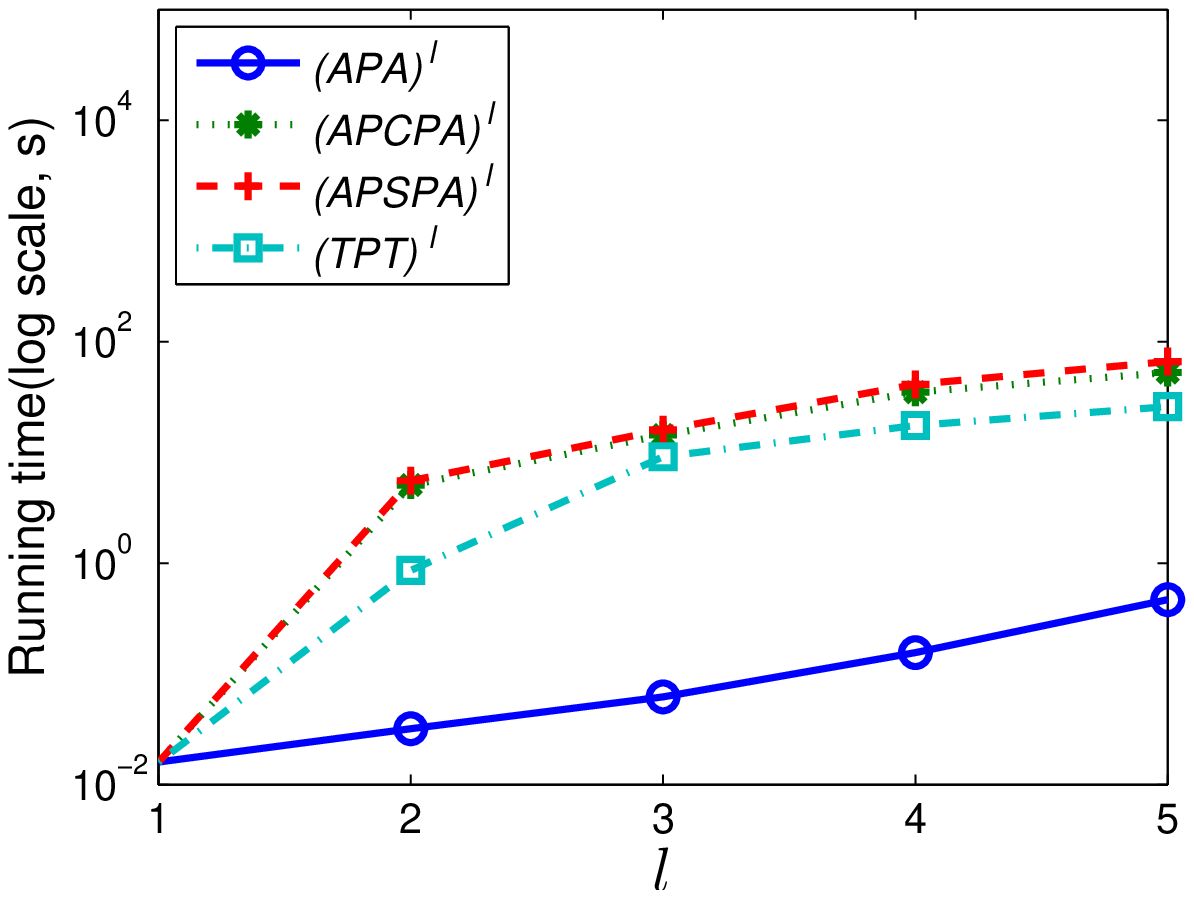}}
    \subfigure[Running time in REL]
    {\includegraphics[scale=0.3]{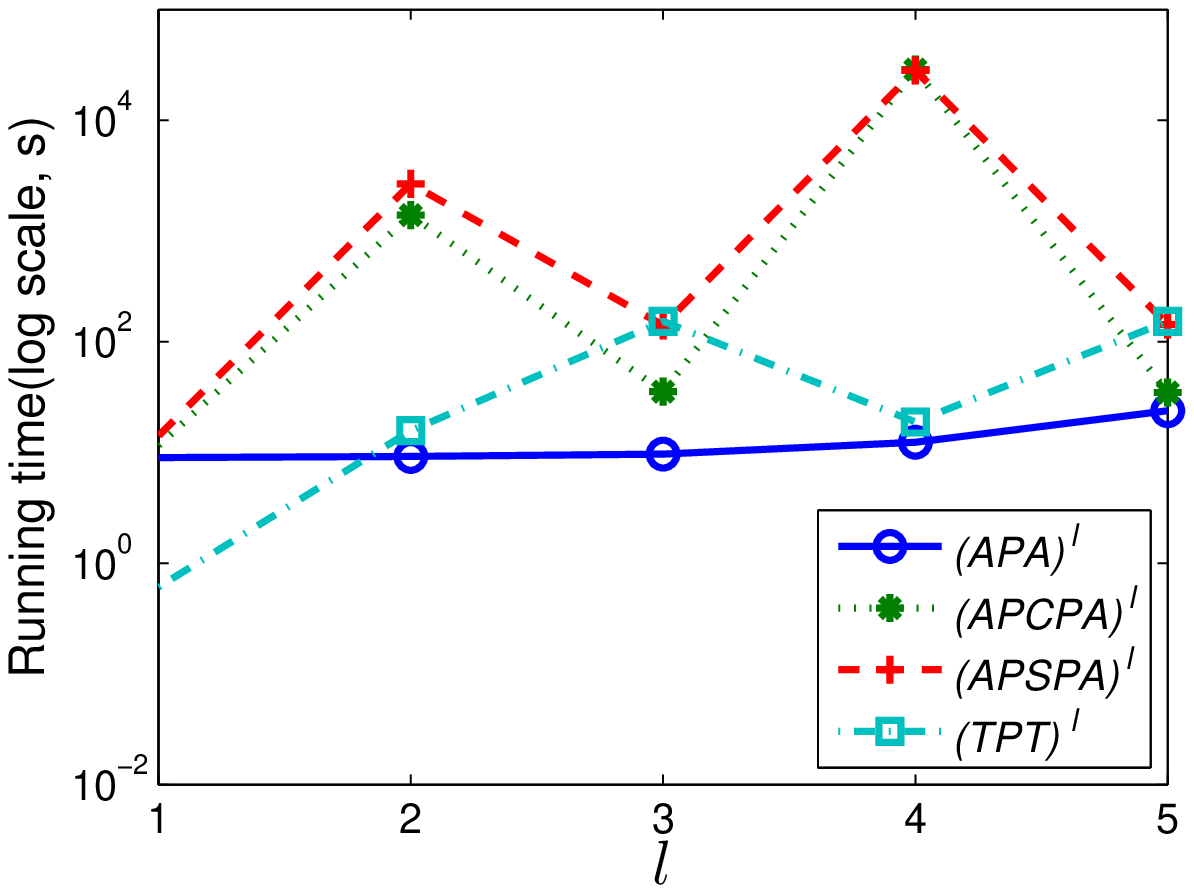}}
    \subfigure[Time ratio in MUL]
    {\includegraphics[scale=0.31]{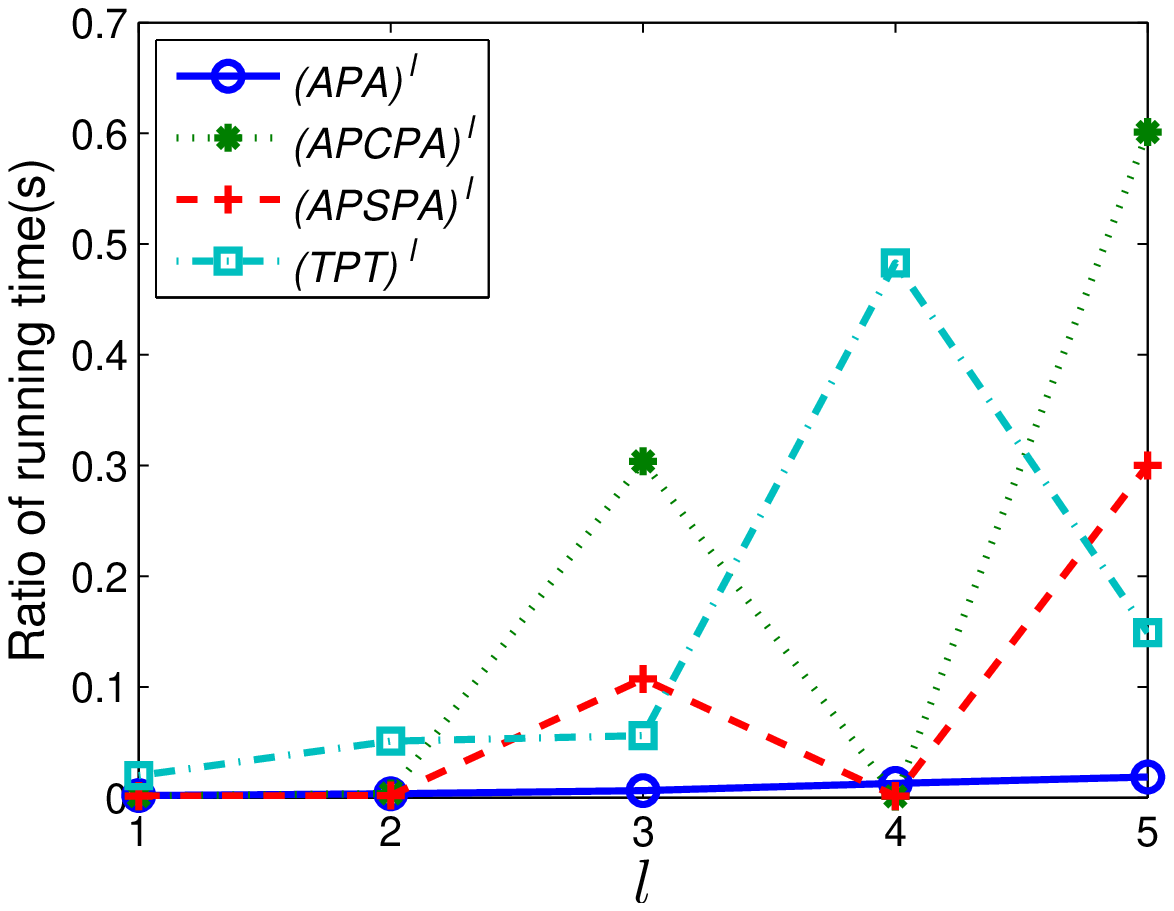}}
    \subfigure[Time ratio in REL]
    {\includegraphics[scale=0.31]{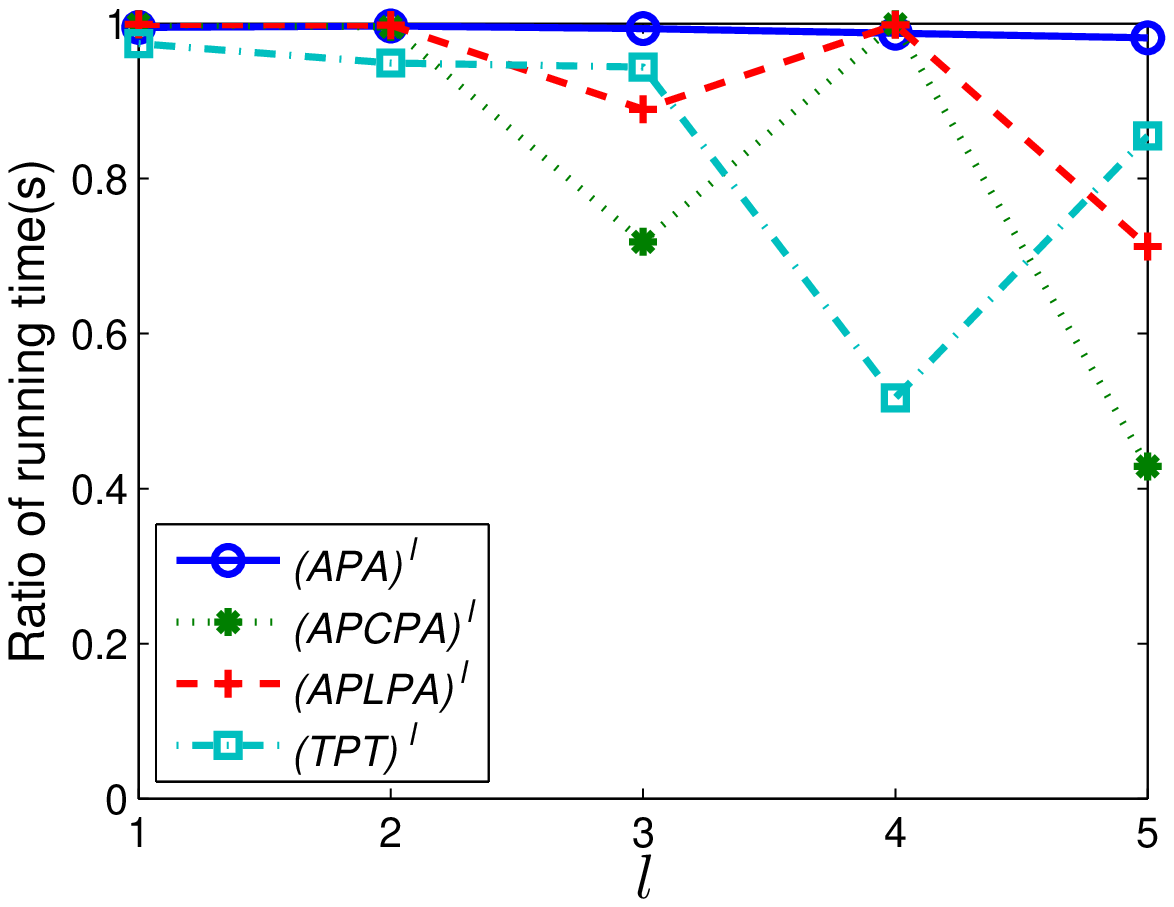}}
    \end{center}
    \caption{Running time of different parts of HeteSim. MUL and REL represent the two components of HeteSim computation (i.e., matrix multiplication and relevance computation), respectively.}\label{PartRunTime}
\end{figure*}

\subsection{Computation Characteristics of HeteSim}

The computation of HeteSim includes two
phases: matrix multiplication (denoted as MUL, i.e., the computation
of $PM_{\mathcal{P}_L}$ and $PM_{{\mathcal{P}_R}^{-1}}$), relevance
computation (denoted as REL, i.e., the computation of
$PM_{\mathcal{P}_L}*PM_{{\mathcal{P}_R}^{-1}}$ and normalization).
In order to analyze the computation characteristics of HeteSim, we
do experiments to observe the running time of these two phases on
different paths with varying path lengths.

Based on the ACM dataset (see Section V.A), we select four paths
with varying length ($l$): ${(APA)}^l$, ${(APCPA)}^l$,
${(APSPA)}^l$, and ${(TPT)}^l$. $l$ means times of path repetition and ranges from 1 to 5. We record
the running time of different phases of HeteSim based on these paths
as shown in Fig. \ref{PartRunTime}. We first observe MUL's running
time in Fig. \ref{PartRunTime}(a). Different paths
have different running time. With the increment of path
length, the running time of matrix multiplication persistently
increase, since more matrix need to be multiplied. Then we consider the
running time of REL phase in Fig. \ref{PartRunTime}(b). Besides
the same observation with Fig. \ref{PartRunTime}(a), the running
time of REL is greatly affected by the length $l$. That is, the
running time of REL significantly increases for ${(APCPA)}^l$ and
${(APSPA)}^l$ when $l$ is 2 and 4.
Let's take the ${(APCPA)}^l$ as an example to analyze the reason. When
$l$ is 1, 3, and 5, the source and target nodes will meet at the
middle node $C$ along the ${(APCPA)}^l$ path, so the relevance
calculation is $PM_{AC}\times PM_{CA}$. However, the relevance calculation
is $PM_{AA}\times PM_{AA}$ when $l$ being 2 and 4. Since the dimension of
$A$ is much larger than that of $C$, the running time of $PM_{AA}\times PM_{AA}$
is much longer than that of $PM_{AC}\times PM_{AC}$. The similar reason makes ${(TPT)}^l$ have the opposite fluctuation. In addition, the time spent in REL does not grow any longer when the matrix become a dense one. So its increase ratio gradually
decreases. For the ${(APA)}^l$ path, the dimension of $A$ and $P$
are close (\# A 17K and \# P 12K), so its running time has no
distinct difference for different path lengths. In addition, the
reachable probability matrix always keeps sparse, which makes the
running time of ${(APA)}^l$ smaller than that of other paths.

Fig. \ref{PartRunTime}(c) and (d) show the ratio of running time in
these two phases over total running time. On one hand, it illustrates that the REL phase
dominates the running time of HeteSim. On the other hand, the ratio
of MUL consistently increases with the increment of
path length. From these experiments, we can summarize two
characteristics of HeteSim computation. (1) The relevance computation
is the main time-consuming phase. It implies that the speedup of matrix
multiplication may not significantly reduce HeteSim's running time,
although this kind of strategies is widely used in accelerating SimRank
\cite{JW02} and PCWR \cite{LC10a}. (2) The dimension and sparsity of matrix greatly affect the efficiency of HeteSim.

\subsection{Quick Computation Strategies}
Although we cannot reduce the running time of relevance computation phase directly,
we can accelerate the computation of HeteSim through adjusting matrix dimension and keeping matrix sparse. Based on above idea, we design the following four strategies.

\subsubsection{Dynamic Programming Strategy}
The matrix multiplication obeys the associative property. Moreover, different computation sequences have different
time complexities. The Dynamic Programming
strategy (DP) changes the sequence of matrix multiplication with the
associative property. The basic idea of DP is to assign
low-dimensioned matrix with the high computation priority. For a
path $\mathcal{P}=R_1\circ R_2\circ \cdots \circ R_{l}$, the
expected minimal computation complexity of HeteSim
can be calculated by the following equation and the computation
sequence is recorded by $i$.

\begin{equation}
\small
\begin{split}
Com(R_1 \cdots R_{l}) =
&\left\{ \begin{array}{lll}
0 & \textrm{$l=1$}\\
|R_1.S|\times|R_1.T|\times|R_2.T| & \textrm{$l=2$}\\
\arg\underset{i}{\min}\{Com(R_1\cdots R_i)+Com(R_{i+1}\cdots R_{l})+|R_1.S|\times|R_i.T|\times|R_l.T|\} &\textrm{$l>2$}\\
\end{array} \right.
\end{split}
\end{equation}
The above equation can be easily solved by dynamic programming
method with the $O(l^2)$ complexity. The running time can be
omitted, since $l$ is much smaller than the matrix dimension.

There may be many duplicate sub-paths in the relevance path.
Obviously, these reduplicative sub-paths only need to be computed
once. For example, the result of \emph{APTPA} can be obtained by computing the matrix $APT$ once. During the matrix multiplication, the DP strategy reserves the computation
sequences of matrices and corresponding results. For a new
computation sequence, if it has been computed before, the corresponding result can be
employed directly. So the reuse strategy further accelerates the matrix multiplication.
Note that the DP strategy only accelerates the MUL phase (i.e.,
matrix multiplication) and it does not change relevance result, so the DP
is a information-lossless strategy.

\subsubsection{Truncation Strategy}
The truncation strategy is based on the hypothesis that removing the
probability on those less important nodes would not significantly
degrade the performance, which have been proved by many researches
\cite{LC10a,SC07}. One advantage of this strategy is to keep matrix
sparse. The sparse matrix greatly reduces the amount of space and
time consumption. The basic idea of truncation strategy is to add a
truncation step at each step of random walk. In the truncation step,
the relevance value is set with 0 for those nodes when their
relevance values are smaller than a threshold $\varepsilon$. A
static threshold is usually used in many methods (e.g., ref. \cite{LC10a}).
However, it has the following disadvantage: it may truncate nothing
for matrix whose elements all have high probability and it may truncate most nodes for matrix whose elements all have low probability. Since we usually pay close attention to the top $k$
objects in query task, the threshold $\varepsilon$ can be set as the
top $k$ relevance value for each search object. For a similarity matrix with size $M\times L$, the $k$ can be dynamicly adjusted as follows.
\begin{displaymath}\small
k = \left\{ \begin{array}{ll}
L & \textrm{if $L\leq W$}\\
\lfloor {(L-W)}^{\beta}\rfloor+W (\beta\in[0,1]) & \textrm{others}\\
\end{array} \right.
\end{displaymath}
where $W$ is the number of top objects, decided by users. The basic idea of dynamic adjustment is that the $k$ slowly increases for super object type (i.e., $L$ is large). The $W$ and $\beta$ determine the truncation
level. The larger $W$ or $\beta$ will cause the larger $k$, which
means a denser matrix. It is expensive to determine the top $k$
relevance value for each object, so we can estimate the value by the top $kM$
value for the whole matrix. Furtherly, the top $kM$
value can be approximated by the sample data with ratio $\gamma$ from the raw
matrix. The larger $\gamma$ leads to
more accurate approximation with longer running time. In summary,
the truncation strategy is an information-loss strategy, which keeps
matrix sparse with small sacrifice on accuracy. In addition, it
needs additional time to estimate the threshold $\varepsilon$.

\subsubsection{Hybrid Strategy}
As discussed above, the DP strategy can accelerate the MUL phase and
the truncation strategy can indirectly speed up the REL phase by keeping sparse matrix. So a hybrid strategy can be designed to combine these two strategies. For the MUL phase, the DP strategy is
applied. After obtaining the $PM_{\mathcal{P}_L}$ and
$PM_{{\mathcal{P}_R}^{-1}}$, the truncation strategy is added.
Different from the above truncation strategy, the hybrid strategy
only truncates the $PM_{\mathcal{P}_L}$ and
$PM_{{\mathcal{P}_R}^{-1}}$. The hybrid strategy utilizes the
benefits of DP and truncation strategies. It is also an
information-loss strategy, since the truncation strategy is employed.

\subsubsection{Monte Carlo Strategy}
Monte Carlo method (MC) is a class of computational algorithms that estimate results through repeating random sampling. It has been applied to compute approximate values of matrix multiplication. Fogaras et al. \cite{FRCS05} applied a Monte Carlo algorithm to compute
approximate personalized PageRank. Recently, Ni et al. \cite{LC10a} tested the effectiveness of the Monte Carlo sampling strategy in the context of path-constrained random walk models.

In this study, we applied the MC strategy to estimate the value of $PM_{\mathcal{P}_L}$ and
$PM_{{\mathcal{P}_R}^{-1}}$. The value of $PM_{\mathcal{P}}(a,b)$ can be approximated by the normalized count of the number of times that the walkers visit the node
$b$ from $a$ along the path $\mathcal{P}$.
\begin{displaymath}\small
PM_{\mathcal{P}}(a,b)=\frac{\# times\ the \ walkers\  visit\  b\  along\  \mathcal{P}}{\# walkers \ from \ a}
\end{displaymath}

The number of walkers from $a$ (i.e., $K$) controls the accuracy and amount of computation. The larger $K$ will achieve more accurate estimation with more time cost. An advantage of the MC strategy is that its running time is not affected by the dimension and sparsity of matrix. However, the high-dimension matrix needs larger $K$ for high accuracy. As a sampling method, the MC is also an information-loss strategy.

\begin{figure*}[t]
    \begin{center}
    \subfigure[Time on ${(APA)}^l$]
    {\includegraphics[scale=0.3]{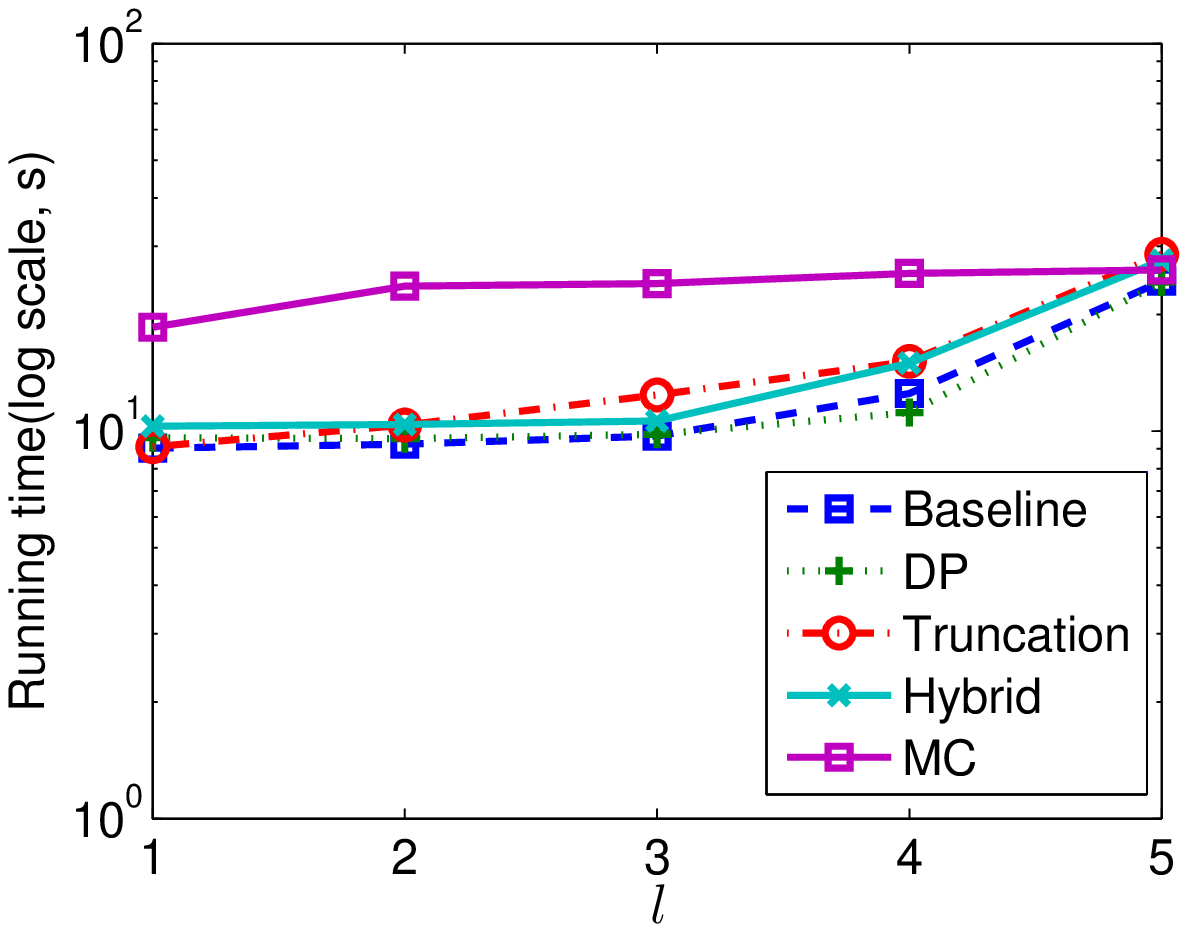}}
    \subfigure[Time on ${(APCPA)}^l$]
    {\includegraphics[scale=0.3]{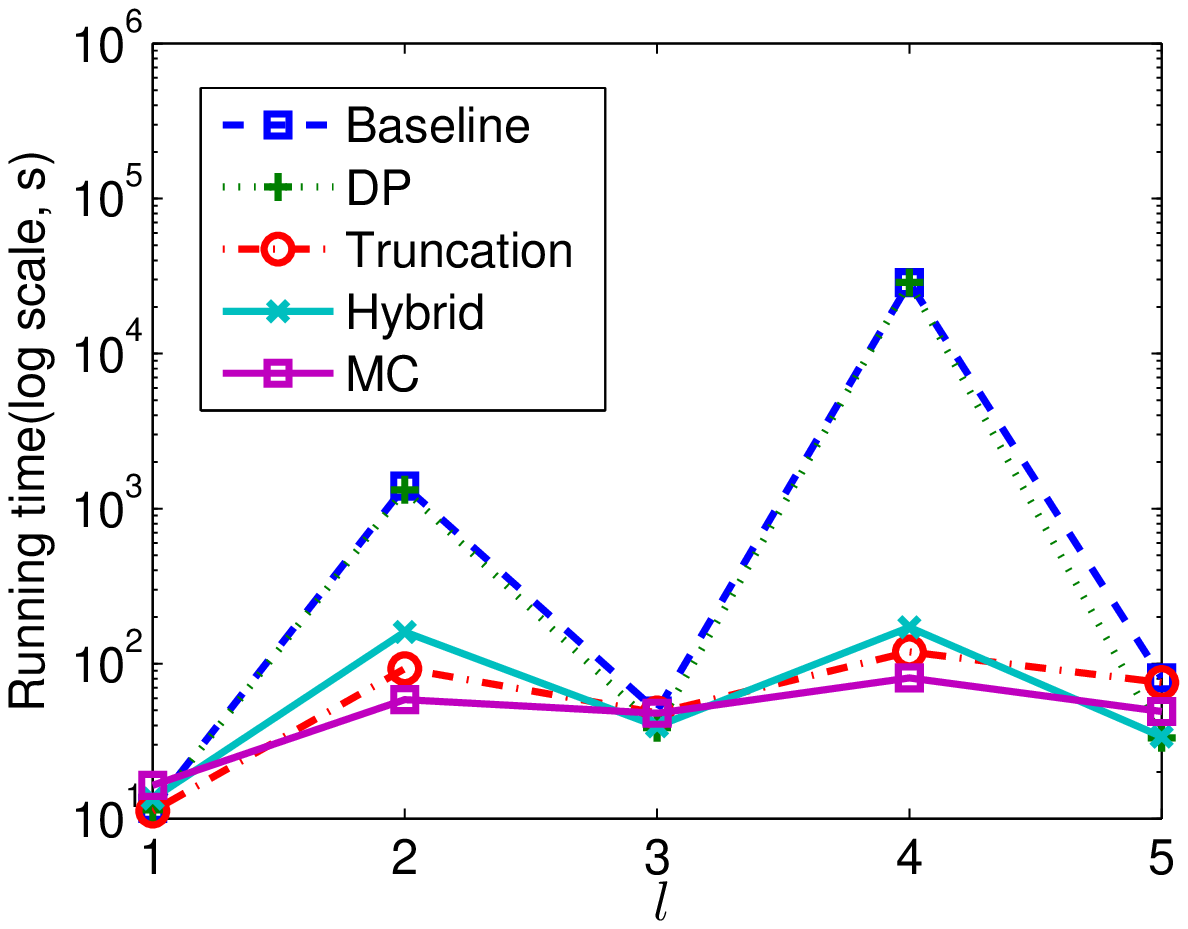}}
    \subfigure[Time on ${(APSPA)}^l$]
    {\includegraphics[scale=0.3]{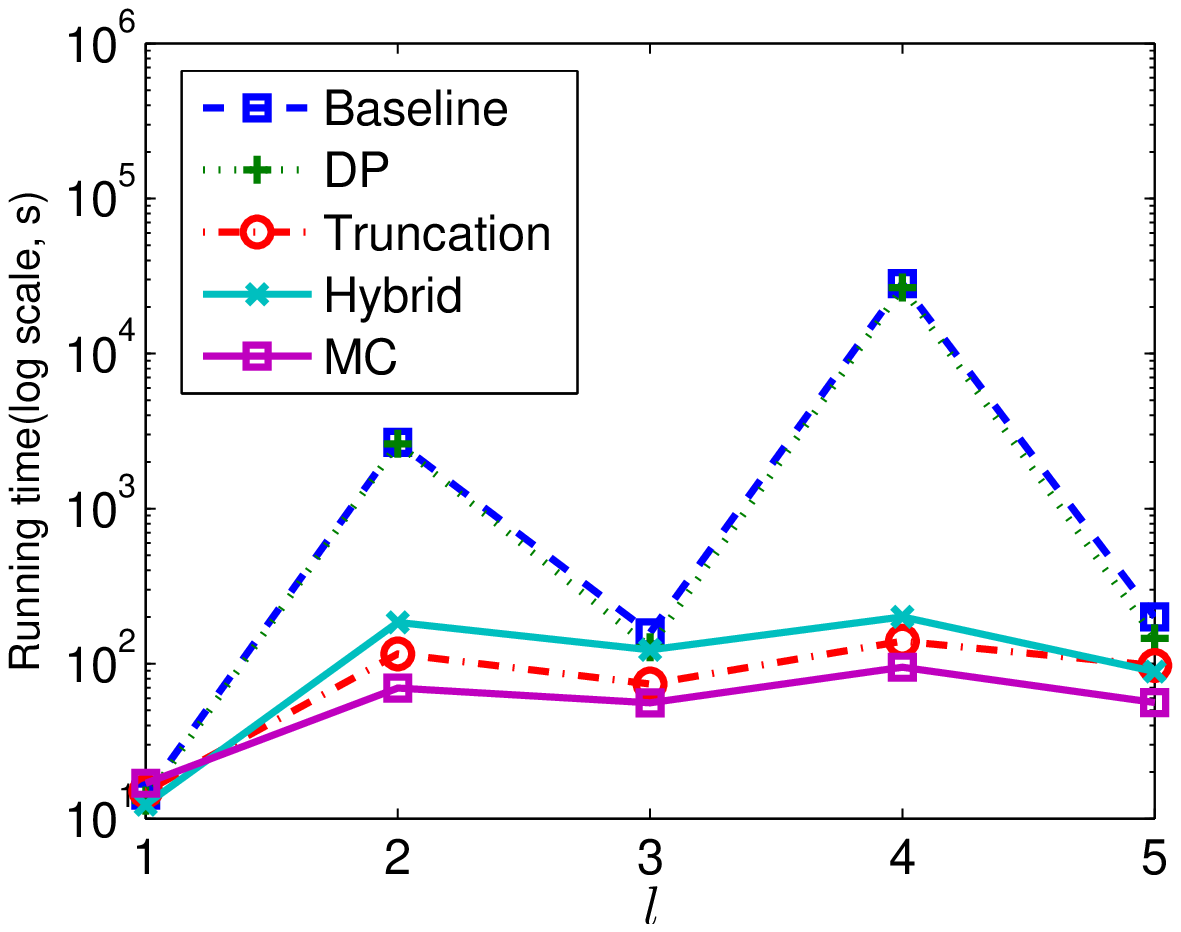}}
    \subfigure[Time on ${(TPT)}^l$]
    {\includegraphics[scale=0.3]{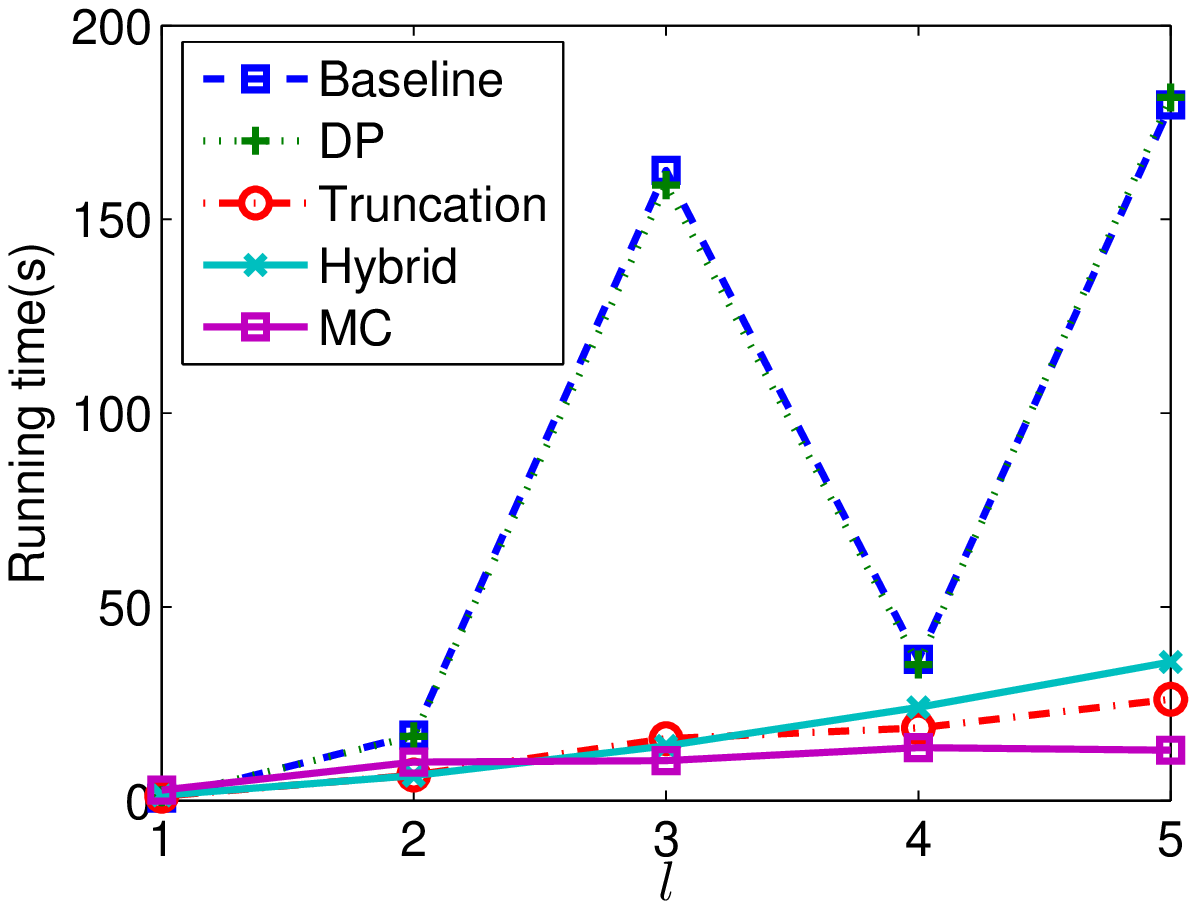}}

    \subfigure[Recall on ${(APA)}^l$]
    {\includegraphics[scale=0.3]{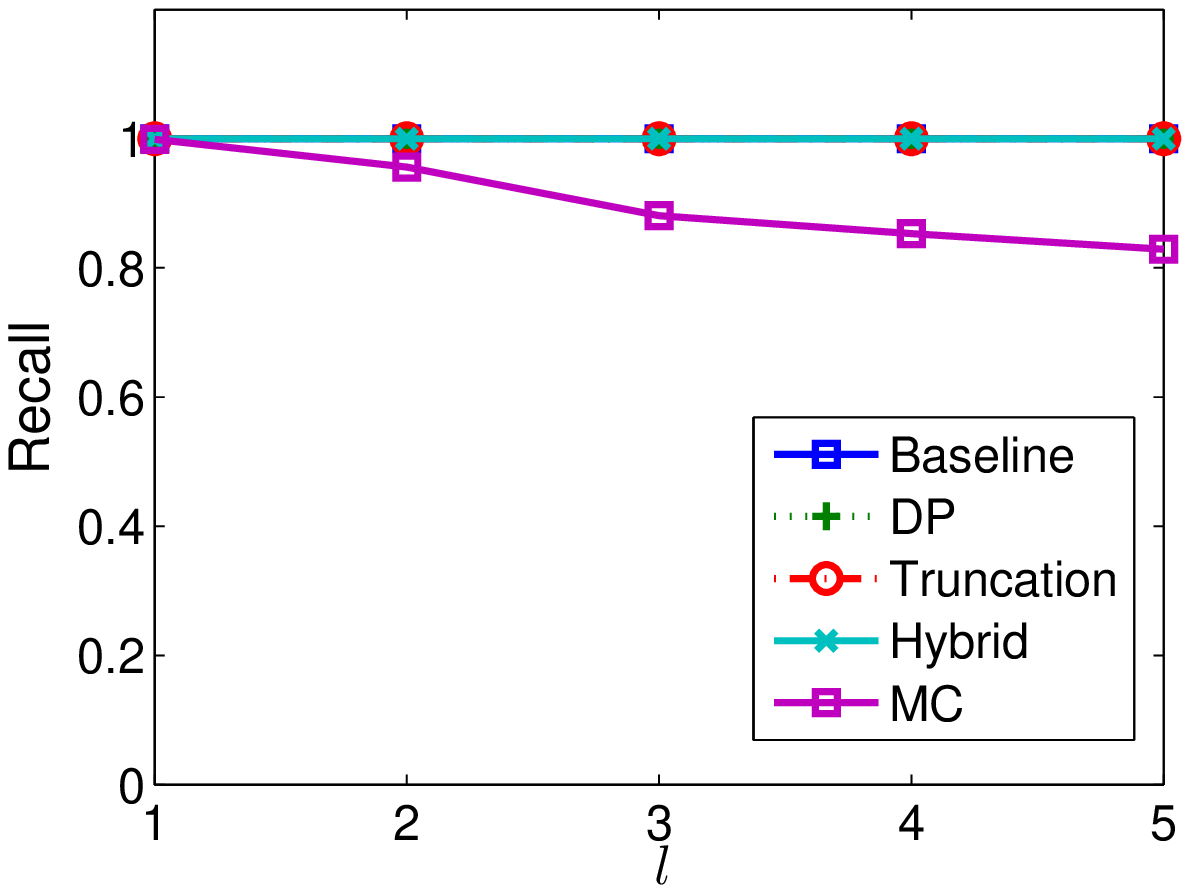}}
    \subfigure[Recall on ${(APCPA)}^l$]
    {\includegraphics[scale=0.3]{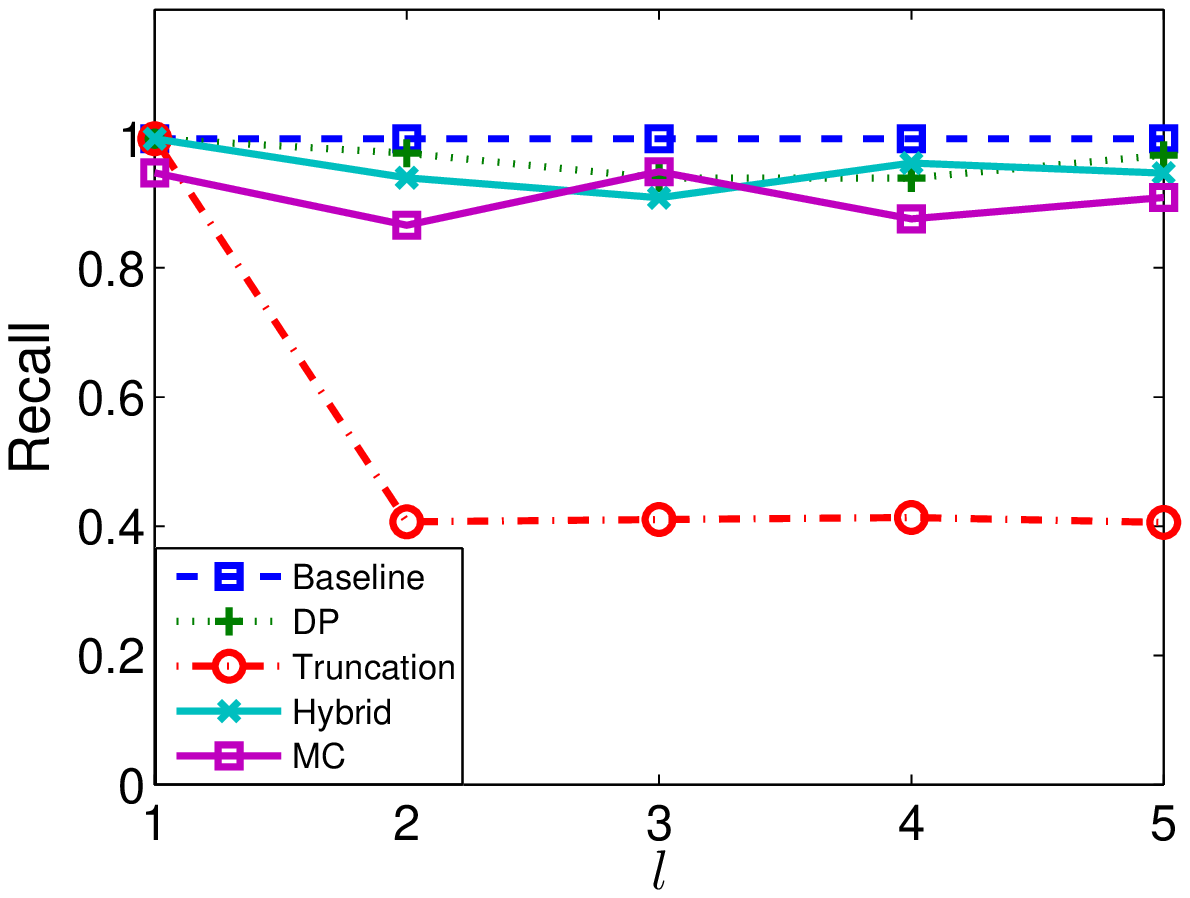}}
    \subfigure[Recall on ${(APSPA)}^l$]
    {\includegraphics[scale=0.3]{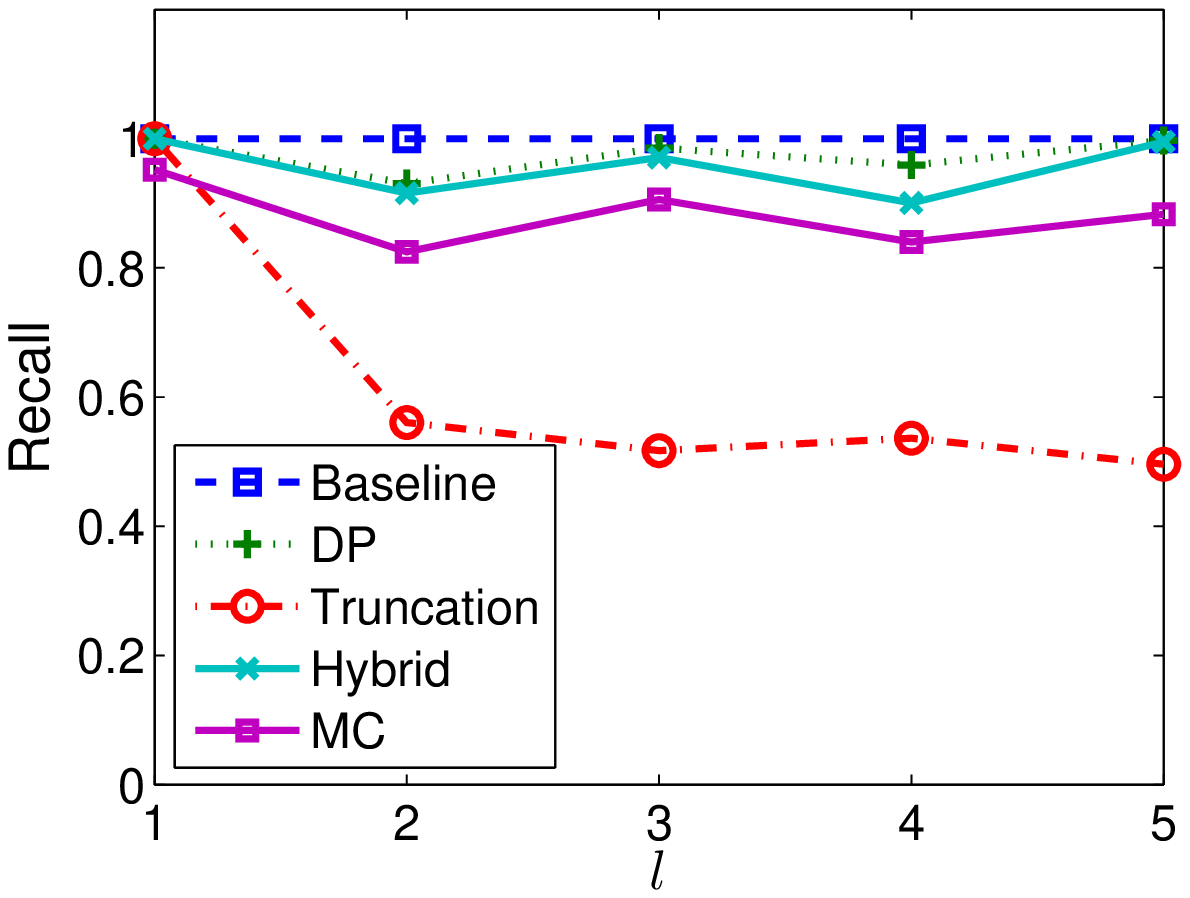}}
    \subfigure[Recall on ${(TPT)}^l$]
    {\includegraphics[scale=0.3]{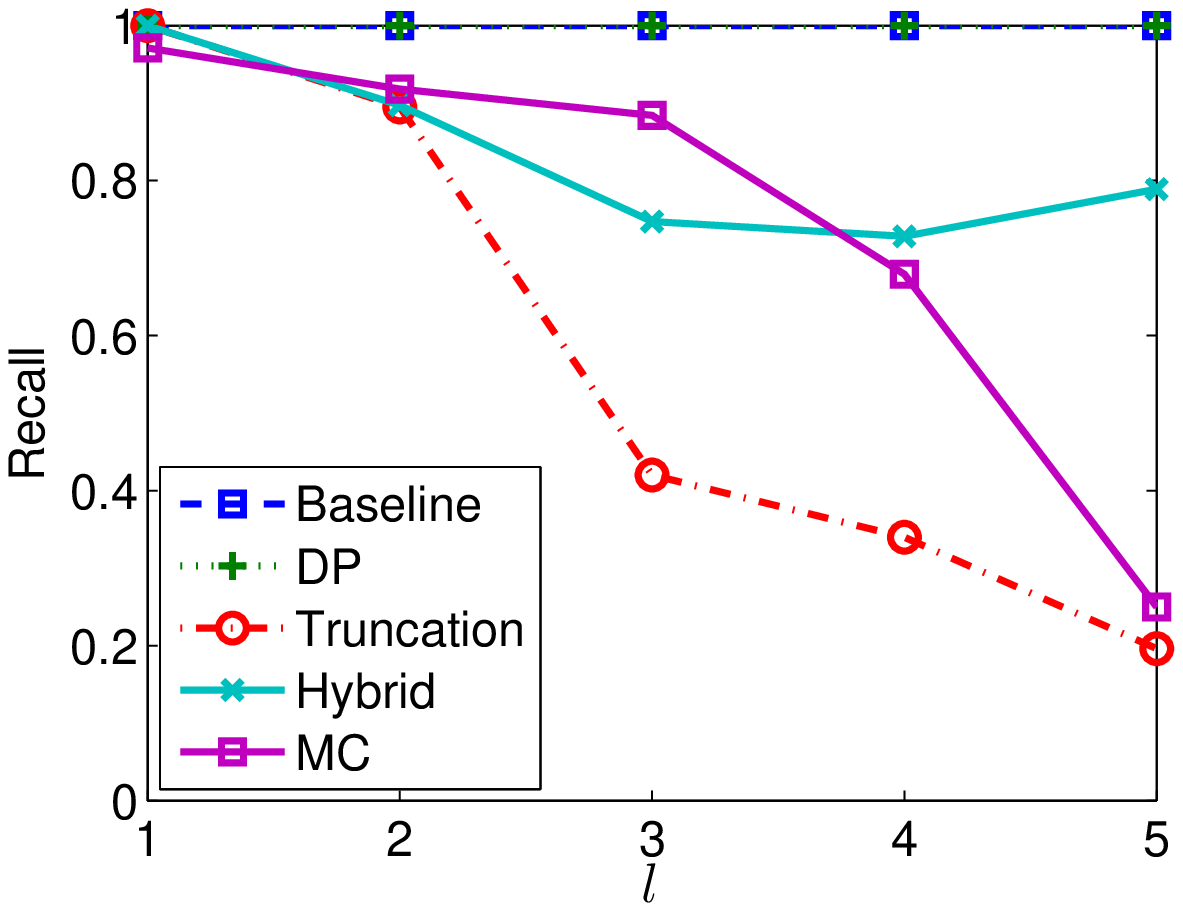}}
    \end{center}
    \caption{Running time and accuracy of computing HeteSim based on different strategies and paths. }\label{TimeAcc}
\end{figure*}

\begin{figure}[t]
    \begin{center}
    \subfigure[MUL phase]
    {\includegraphics[scale=0.3]{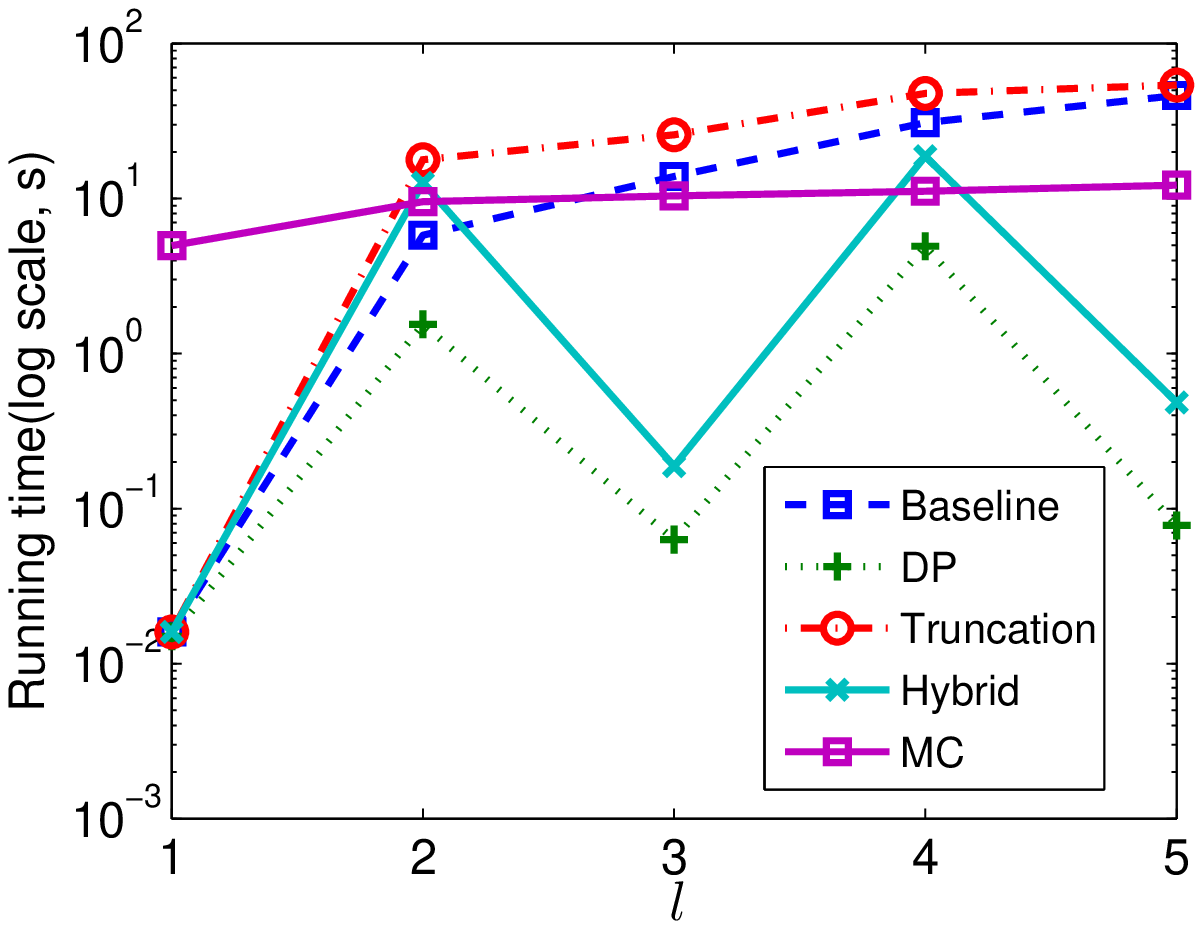}}
    \subfigure[REL phase]
    {\includegraphics[scale=0.3]{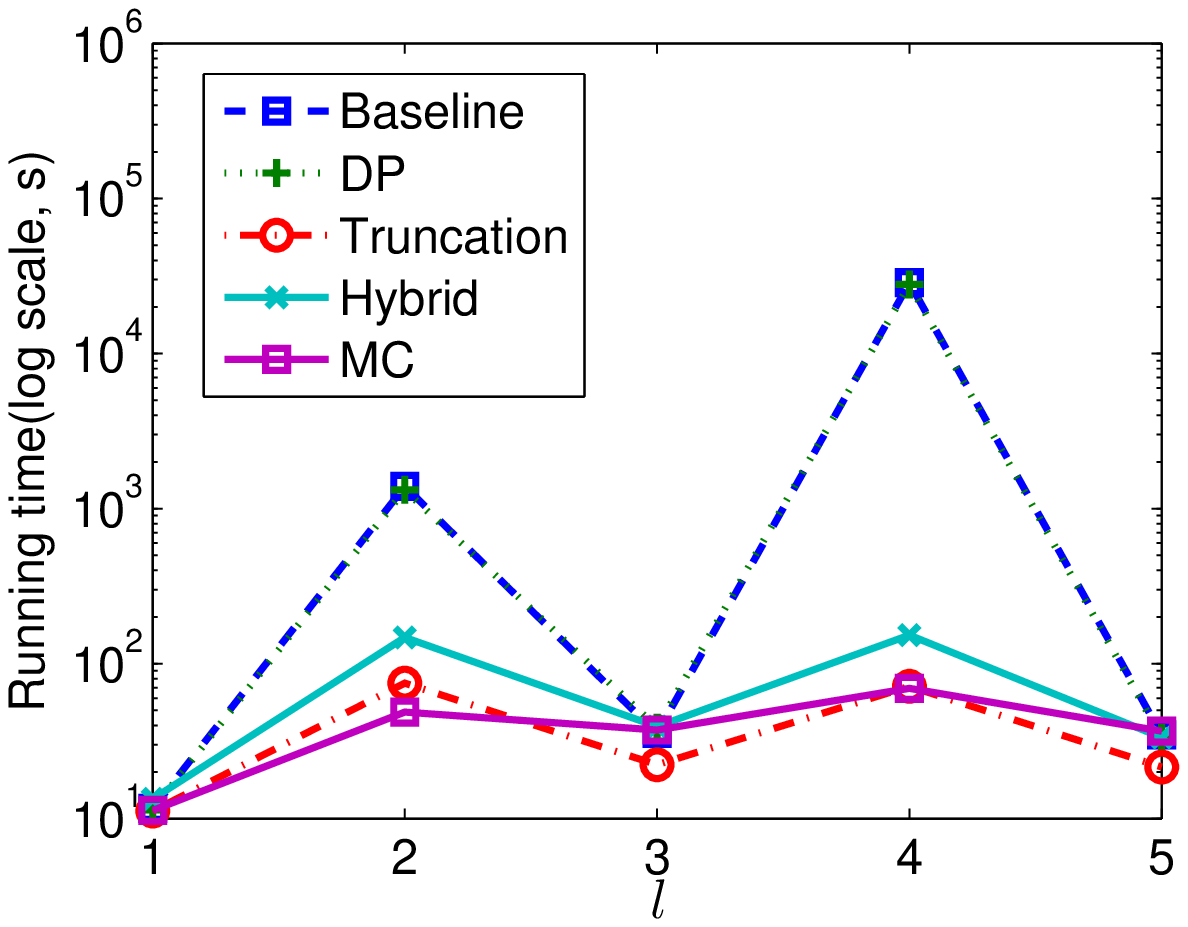}}
    \end{center}
    \caption{Different phase's running time of computing HeteSim based on ${(APCPA)}^l$ path.}\label{DiffPhase}
\end{figure}

\subsection{Quick Computation Experiments}
We validate the efficiency and effectiveness of quick computation
strategies on the ACM dataset. The four paths
are used: ${(APA)}^l$, ${(APCPA)}^l$, ${(APSPA)}^l$, and
${(TPT)}^l$. $l$ means times of path repetition and ranges from 1 to 5. Four quick computation
strategies and the original method (i.e., baseline) are employed.
The parameters in truncation process are set as follows: the
number of top objects $W$ is 200, $\beta$ is 0.5, and $\gamma$ is
0.005. The number of walkers (i.e., $K$) in MC strategy is 500. The running time and accuracy of all strategies are recorded. In the accuracy evaluation, the relevance matrix obtained by the
original method are regarded as the baseline. The accuracy is the
$recall$ criterion on the top 100 objects obtained by each strategy.
All experiments are conducted on machines with Intel Xeon 8-Core
CPUs of 2.13 GHz and 64 GB RAM.

Fig. \ref{TimeAcc} shows the running time and accuracy of four
strategies on different paths. The running time of these strategies
are illustrated in Fig. \ref{TimeAcc} (a)-(d). We can observe that
the DP strategy almost has the same running time with the baseline.
It only speeds up the HeteSim computation when the MUL
phase dominates the whole running time (e.g., ${(APCPA)}^5$ and
${(APSPA)}^5$). It is not the case for the truncation and hybrid strategies, which significantly accelerate the HeteSim computation and have a close speedup ratio on most conditions. Except the $APA$ path, the MC strategy has the highest speedup ratio among all four strategies on most conditions. Then, let's observe their accuracy from Fig. \ref{TimeAcc} (e)-(h). The accuracy of the DP strategy is always
close to 1. The hybrid strategy achieves the second performances for most paths. The accuracy of the MC strategy is also high for most paths, while it fluctuates on different paths. Obviously, the truncation strategy has the lowest accuracy on most conditions.

As we have noted, the DP is an information-lossless strategy and it
only speeds up the MUL phase. Moreover, the MUL phase is not the
main time-consuming part for most paths. So the DP strategy
trivially accelerates HeteSim with the accuracy close to 1. The
truncation strategy is an information-loss strategy to keep matrix
sparse, so it can effectively accelerate HeteSim. That is the reason why
the truncation strategy has the high speedup ratio but low accuracy. The
hybrid strategy combines the DP and truncation strategy. So it has
a close speedup ratio to the truncation strategy. The hybrid
strategy only does truncation on the last step of random walk, which makes it less
information-loss. It explains that its accuracy is higher than the truncation
strategy. As we know, the essence of the MC strategy is repeatedly random sampling. In order to achieve high accuracy, more walkers (i.e., larger $K$) are needed for high-dimension or sparse matrix. In our experiments, the fixed walkers (i.e., $K$ is 500) makes the MC strategy the poor accuracy on some conditions. For example, in Fig. \ref{TimeAcc} (h), the relevance calculation is $PM_{TP}\times PM_{TP}$ for ${(TPT)}^5$. The high dimension of $P$ and even distribution result in the low accuracy of MC strategy.

In order to clearly illustrate the effect of these
strategies on two phases of HeteSim computation, a typical running-time example on
${(APCPA)}^l$ is shown in Fig. \ref{DiffPhase}. It is clear that
the DP strategy greatly accelerates the MUL phase indeed, but it has no
effect on the REL phase. On the contrary, the truncation strategy is
slower than the baseline on the MUL phase, due to sparse matrix and
additional time spent in estimating the threshold. However, the truncation strategy greatly accelerates the REL phase because the sparse matrix is kept. Compared to the truncation strategy, the MC strategy not only accelerates the REL phase but also benefits the MUL phase on dense matrix.

According to the analysis above, these strategies are suitable for
different paths and scenarios. For very sparse matrix (e.g.,
${(APA)}^l$) and low-dimension matrix (e.g., ${(APCPA)}^3$), all
strategies cannot significantly improve efficiency. However, in
these conditions, the HeteSim can be quickly computed without applying any quick computation strategies. For those dense (e.g., ${(APCPA)}^4$) and high-dimension matrix (e.g., ${(APSPA)}^4$) which has huge computation overhead, the truncation, hybrid, and MC strategies can effectively improve the
HeteSim's efficiency. Particularly, the speedup of the hybrid and MC
strategies are up to 100 with little loss in accuracy. If the MUL phase
is the main time-consuming part for a path, the DP strategy can also
speed up HeteSim greatly without loss in accuracy. The MC strategy has very high efficiency, but its accuracy may degrade for high-dimension matrix. So the appropriate
$K$ needs to be set through balancing the efficiency and effectiveness.

\section{Conclusion}
In this paper, we study the relevance search problem which measures
the relatedness of heterogeneous objects (including same-typed or
different-typed objects) in heterogeneous networks. We propose a
general relevance measure, called HeteSim. As a path-constraint
measure, HeteSim can measure the relatedness of same-typed and
different-typed objects in a uniform framework. In addition, HeteSim
is a semi-metric measure, which can be used in many applications.
Extensive experiments validate the effectiveness and efficiency of HeteSim on
evaluating the relatedness of heterogeneous objects.

There are some interesting directions for future work. Firstly, more methods can be explored to measure the relatedness of heterogeneous objects, such as path count and RW strategies. Secondly, since the quick computation strategies proposed in this paper are all in-memory methods, the parallel computation methods of HeteSim can be an interesting topic to explore. Last but not least, the problem on how to choose and weight different meta paths are also important issues for heterogeneous networks.

\balance



%

\newpage

\bibliographystyle{IEEEtran}


\newpage
\appendices
\section{Proof of Properties}

\textbf{Proof of Property 1}.
According to Definition 6, for each
relation instance $a\rightarrow b$ in relation $R=AB$ ($a\in A$ and
$b\in B$), add an object $e$ ($e\in E$) between $a$ and $b$, and let
$w_{ae}=w_{eb}=\sqrt{w_{ab}}$ where $w$ means the weight of relation
instances. Note that for adjacent matrix,
$w_{ae}=w_{eb}=\sqrt{w_{ab}}$=1. Since $a$ and $b$ only meet on $e$,
so $R_O(a,:)*R_I(:,b)=w_{ae}*w_{eb}=w_{ab}=R(a,b)$. So $R=R_O\circ
R_I$. Since the process is unique, the decomposition is unique.

\textbf{Proof of Property 2}.
According to Definition 8, $U_{AB}$ is
the normalized matrix of the transition probability matrix $W_{AB}$
along the row vector, which is also the transposition of the
normalized matrix of $W_{BA}$ along the column vector (i.e.,
$V_{BA}$). So $U_{AB}=V'_{BA}$. Similarly, $V_{AB}=U'_{BA}$.

\textbf{Proof of Property 3}. According to Definition 5,
$\mathcal{P}=\mathcal{P}_L\mathcal{P}_R$ and
$\mathcal{P}^{-1}={\mathcal{P}_R}^{-1}{\mathcal{P}_L}^{-1}$.
According to Equation 6,
\begin{equation}\label{}
\begin{split}
&HeteSim(a,b|\mathcal{P})=\frac{PM_{\mathcal{P}_L}(a,:)PM'_{{\mathcal{P}_R}^{-1}}(b,:)}{\sqrt{\|PM_{\mathcal{P}_L}(a,:)\|\|PM_{{\mathcal{P}_R}^{-1}}(b,:)\|}}\\
&HeteSim(b,a|\mathcal{P}^{-1})=\frac{PM_{{\mathcal{P}_R}^{-1}}(b,:)PM'_{\mathcal{P}_L}(a,:)}{\sqrt{\|PM_{{\mathcal{P}_R}^{-1}}(b,:)\|\|PM_{\mathcal{P}_L}(a,:)\|}}
\end{split}
\end{equation}
so $HeteSim(a,b|\mathcal{P})=HeteSim(b,a|\mathcal{P}^{-1})$.

\textbf{Proof of Property 4}.
According to Equation 6,
$HeteSim(a,b|\mathcal{P})=cos(PM_{\mathcal{P}_L}(a,:),
PM_{{\mathcal{P}_R}^{-1}}(b,:))\in [0,1]$. If and only if
$PM_{\mathcal{P}_L}(a,:)$ is equal to
$PM_{{\mathcal{P}_R}^{-1}}(b,:)$, $cos(PM_{\mathcal{P}_L}(a,:),
PM_{{\mathcal{P}_R}^{-1}}(b,:))=1$, so $HeteSim(a,b|\mathcal{P})=1$.

\textbf{Proof of Property 5}. It is obvious that
$SimRank_0(a_1,a_2)=HeteSim(a_1,a_2|I)$ and \\
$SimRank_0(b_1,b_2)=HeteSim(b_1,b_2|I)$. Here $SimRank_i$ means
SimRank value after $i$ hop. Let's consider the 1st hop condition.
\begin{equation}
\begin{split}
 &SimRank_1(a_1,a_2)\\
 &=\frac{1}{|O(a_1)||O(a_2)|}\sum_{i=1}^{|O(a_1)|}\sum_{j=1}^{|O(a_2)|}SimRank_0(O_i(a_1),O_j(a_2))\\
 &=\frac{1}{|O(a_1)||O(a_2)|}\sum_{i=1}^{|O(a_1)|}\sum_{j=1}^{|O(a_2)|}SimRank_0(b_i,b_j)\\
 &=\frac{1}{|O(a_1)||O(a_2)|}\sum_{i=1}^{|O(a_1)|}\sum_{j=1}^{|O(a_2)|}HeteSim(b_i,b_j|I)\\
 &=HeteSim(a_1,a_2|RR^{-1})
\end{split}
\end{equation} 
since $O(a_2)=I(a_2|BA)$, $O(a_1)=O(a_1|AB)$ and
$SimRank_0(b_1,b_2)=HeteSim(b_1,b_2|I)$. Similarly,
$SimRank_1(b_1,b_2)=HeteSim(b_1,b_2|R^{-1}R)$.
Suppose it is correct
for $k$-th hop, let's consider the $k+1$ hop.
\begin{equation}
\begin{split}
 &SimRank_{k+1}(a_1,a_2)\\
 &=\frac{1}{|O(a_1)||O(a_2)|}\sum_{i=1}^{|O(a_1)|}\sum_{j=1}^{|O(a_2)|}SimRank_k(O_i(a_1),O_j(a_2))\\
 &=\frac{1}{|O(a_1)||O(a_2)|}\sum_{i=1}^{|O(a_1)|}\sum_{j=1}^{|O(a_2)|}SimRank_k(b_i,b_j)\\
 &=\frac{1}{|O(a_1)||O(a_2)|}\sum_{i=1}^{|O(a_1)|}\sum_{j=1}^{|O(a_2)|}HeteSim(b_i,b_j|(R^{-1}R)^k)\\
 &=HeteSim(a_1,a_2|R(R^{-1}R)^kR^{-1})\\
 &=HeteSim(a_1,a_2|(RR^{-1})^{k+1})\\
\end{split}
\end{equation} 
Similarly,
$SimRank_{k+1}(b_1,b_2)=HeteSim(b_1,b_2|(R^{-1}R)^{k+1})$ So
\begin{equation}
\begin{split}
 SimRank(a_1,a_2)
 &=\underset{n\rightsquigarrow \infty}{lim}\sum_{k=1}^{n}{SimRank_k(a_1,a_2)}\\
 &=\underset{n\rightsquigarrow
 \infty}{lim}\sum_{k=1}^{n}{HeteSim(a_1,a_2|(RR^{-1})^k)}
\end{split}
\end{equation} 
Similarly,
\begin{equation}
\begin{split}
 SimRank(b_1,b_2)
 &=\underset{n\rightsquigarrow
 \infty}{lim}\sum_{k=1}^{n}{HeteSim(b_1,b_2|(R^{-1}R)^k)}
\end{split}
\end{equation} 

\section{Automatic Object Profiling}
In this case study, we want to find the profile of KDD
conference. Table \ref{tab:ConfSearch} shows the results on ACM
dataset. The active researchers in the conference can be found by
the $CVPA$ path indicating the relationship of authors publishing
papers in conferences. The top five authors are all well-known
researchers in data mining area. The $CVPAF$ path reveals the
important research affiliations that have published many papers in
KDD, such as CMU, IBM, Yahoo! Research. The results of $CVPS$
illustrate that the topics of KDD are database management (H.2),
pattern recognition (I.5), and so on. The $CVPAPVC$ path measures
the similarity of conferences through their common authors. The
conferences that are most similar to KDD are VLDB, SIGMOD, WWW and
CIKM. It is reasonable, since these conferences all share many
authors whose research areas are data mining and knowledge
management.

\begin{table*}
\centering \caption{Automatic object profiling task on conference
``KDD" on ACM dataset.}\label{tab:ConfSearch} \scriptsize
    \begin{tabular}{|c|c|c|c|c|c|c|c|c|}
        \hline
        Path   &\multicolumn{2}{|c|}{\emph{CVPA}}  & \multicolumn{2}{|c|}{\emph{CVPAF}} &\multicolumn{2}{|c|}{\emph{CVPS}}&\multicolumn{2}{|c|}{\emph{CVPAPVC}}\\
        \hline
        Rank & Authors & Score  & Organization             & Score  & Subjects                                     & Score  & Conf. & Score  \\ \hline
        1    & Christos Faloutsos & 0.1198 & Carnegie Mellon Univ.                 & 0.0824  & H.2 (database management)   & 0.3215 & KDD    & 1      \\
        2    & Heikki Mannila     & 0.1119 & Univ. of Minnesota                    & 0.0814 & I.5 (pattern recognition)   & 0.1650  & VLDB   & 0.2124 \\
        3    & Padhraic Smyth     & 0.1043 & IBM                                        & 0.0761 & I.2 (artificial intelligence)   & 0.1194 & SIGMOD & 0.1535 \\
        4    & Jiawei Han         & 0.1029 & Yahoo! Research                            & 0.0692 & G.3 (prob. and stat.)   & 0.0856 & WWW    & 0.1391 \\
        5    & Vipin Kumar        & 0.0966 & Univ. of California                   & 0.0683 & H.3 (info. storage and retrieval)  & 0.0653 & CIKM   & 0.0943 \\
        \hline
    \end{tabular}
\normalsize
\end{table*}

\section{Evaluation of Relative Importance}
\begin{figure}
    \begin{center}
    {\includegraphics[scale=0.35]{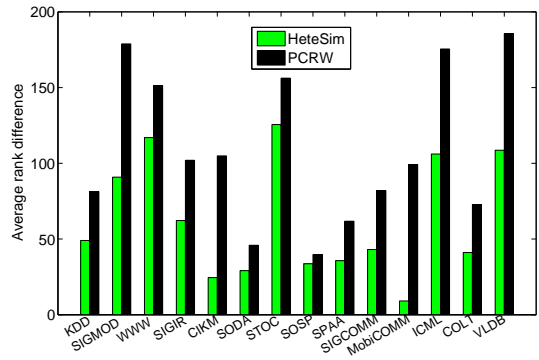}}
    \end{center}
    \caption{The average rank difference of HeteSim and PCRW on 14 conferences of ACM dataset. The lower the better.}\label{RankDiff}
\end{figure}

The relative importance is hard to quantitatively measure. However,
we can roughly measure the relatedness of authors and conferences by
the number of papers that authors publish in conferences, and then
rank the relatedness as their relative importance (i.e., ground
truth). We also compute the relatedness of authors and conferences
based on HeteSim and PCRW, and then rank these values. Through
computing the average rank difference from the ground truth, we can
roughly measure the accuracy of relative importance. For example, C.
Faloutsos is ranked 1st on KDD as ground truth, while an approach
rank him 6th. So the rank difference is 5. Note that, since PCRW has
two rank scores for two different orders, the results are the
average rank differences based on these two different orders. Fig.
\ref{RankDiff} show the average rank difference on the top 200
authors in ground truth on each conference. It is clear that HeteSim
more accurately reveals the relative importance of author-conference
pairs, since their average rank difference is smaller.

\begin{table}
\centering \caption{The top 10 most related authors to ``KDD"
conference under different relevance paths on ACM
dataset.}\label{tab:PathSema} \scriptsize
    \begin{tabular}{|c|c|c|}
        \hline
             &\multicolumn{2}{|c|}{path}\\
        \cline{2-3}
             rank&\emph{CVPA}&\emph{CVPAPA}\\
        \hline
         1   & Christos Faloutsos & Charu C. Aggarwal \\
        \hline
         2   & Heikki Mannila   & Philip Yu \\
         \hline
         3   & Padhraic Smyth &Heikki Mannila \\
        \hline
         4   & Jiawei Han   & Christos Faloutsos  \\
         \hline
         5   & Vipin Kumar  &Jiawei Han  \\
        \hline
         6   & Philip Yu    & Bianca Zadrozny  \\
         \hline
         7   & Eamonn Keogh & Padhraic Smyth \\
        \hline
         8   & Kenji Yamanishi   & Kenji Yamanishi   \\
         \hline
         9   & Mohammed J. Zaki & Inderjit S. Dhillon \\
        \hline
         10  & Charu C. Aggarwal & Vipin Kumar \\
        \hline
    \end{tabular}
\normalsize
\end{table}

\section{Semantic Meaning of Relevance Path}
We know that different paths have different semantic meanings in
heterogeneous networks. Table \ref{tab:PathSema} shows such a case,
which searches the most related authors to KDD conference based on
two different relevance paths. The \emph{CVPA} path means
conferences publishing papers written by authors. It identifies the
most active authors to the conference. The \emph{CVPAPA} path means
conferences publishing papers written by authors' co-authors. It
identities the persons with the most active group of co-authors. In
social network setting, this is like identifying the persons with
the most active group of friends or potential  targets for viral
marketing. At first glance, there are no obvious difference between
the results returned by these two paths. However, the different
ranks of these authors reveal the subtle semantics on the paths. The
\emph{CVPA} path returns authors that have high publication records
in KDD. For example, Christos Faloutsos published the most papers
(32) in KDD. Note that HeteSim does not simply count the number of
paths connecting two objects. It also considers the mutual influence
of two objects. For example, Jiawei Han and Philip Yu published the
second and third highest number of papers in KDD. However, they have
wider research interests and published many papers in many other
conferences, so their relatedness to KDD decrease based on the
\emph{CVPA} path.

By contrast, the \emph{CVPAPA} path emphasizes on the publication
records of the co-authors. The results also reflect this point. For
example, although Charu C. Aggarwal published 13 papers in KDD, not
the highest publication records, he has many co-authors which
include many high-publication-record authors (e.g., Philip Yu and
Jiawei Han), so he is the first author related to KDD based on
\emph{CVPAPA} path. The same thing also happens to other authors.
Taking Bianca Zadrozny for example, she only published 6 papers in
KDD. However, her co-authors also include many
high-publicaton-record authors, such as Philip Yu, Naoki Abe, and
Wei Fan. In all, HeteSim can accurately capture the semantics under
relevance paths.

\begin{IEEEbiographynophoto}{}

\end{IEEEbiographynophoto}




\end{document}